\documentclass[twocolumn]{aastex6}

\usepackage{color}

\newcommand{\sunrise}{\textsc{Sunrise}}
\newcommand{\carcsec}{$\mbox{.\hspace{-0.5ex}}^{\prime\prime}$}

\begin{document}

\title{A new MHD-assisted Stokes inversion technique}

\author{\textsc{
T.~L.~Riethm\"uller,$^{1}$
S.~K.~Solanki,$^{1,2}$
P.~Barthol,$^{1}$
A.~Gandorfer,$^{1}$
L.~Gizon,$^{1,3}$
J.~Hirzberger,$^{1}$
M.~van~Noort,$^{1}$
J.~Blanco~Rodr\'{\i}guez,$^{4}$
J.~C.~Del~Toro~Iniesta,$^{5}$
D.~Orozco~Su\'arez,$^{5}$
W.~Schmidt$^{6}$
V.~Mart\'{\i}nez Pillet,$^{7}$
\& M.~Kn\"olker,$^{8}$
}}
\affil{
$^{1}$Max-Planck-Institut f\"ur Sonnensystemforschung, Justus-von-Liebig-Weg 3, 37077 G\"ottingen, Germany; riethmueller@mps.mpg.de\\
$^{2}$School of Space Research, Kyung Hee University, Yongin, Gyeonggi, 446-701, Republic of Korea\\
$^{3}$Institut f\"ur Astrophysik, Georg-August-Universit\"at G\"ottingen, Friedrich-Hund-Platz 1, 37077 G\"ottingen, Germany\\
$^{4}$Grupo de Astronom\'{\i}a y Ciencias del Espacio, Universidad de Valencia, 46980 Paterna, Valencia, Spain\\
$^{5}$Instituto de Astrof\'{\i}sica de Andaluc\'{\i}a (CSIC), Apartado de Correos 3004, 18080 Granada, Spain\\
$^{6}$Kiepenheuer-Institut f\"ur Sonnenphysik, Sch\"oneckstr. 6, 79104 Freiburg, Germany\\
$^{7}$National Solar Observatory, 3665 Discovery Drive, Boulder, CO 80303, USA\\
$^{8}$High Altitude Observatory, National Center for Atmospheric Research, P.O. Box 3000, Boulder, CO 80307-3000, USA\\
}

\begin{abstract}

We present a new method of Stokes inversion of spectropolarimetric data and evaluate it by taking the example of a
\sunrise{}/IMaX observation. An archive of synthetic Stokes profiles is obtained by the spectral synthesis of
state-of-the-art magnetohydrodynamics (MHD) simulations and a realistic degradation to the level of the observed data. The definition of a merit function
allows the archive to be searched for the synthetic Stokes profiles that match the observed profiles best.
In contrast to traditional Stokes inversion codes, which solve the Unno-Rachkovsky equations
for the polarized radiative transfer numerically and fit the Stokes profiles iteratively, the new technique provides the
full set of atmospheric parameters. This gives us the ability to start an MHD simulation that takes the inversion
result as initial condition. After a relaxation process of half an hour solar time we obtain physically consistent MHD data sets with
a target similar to the observation. The new MHD simulation is used to repeat the method in a second iteration, which further
improves the match between observation and simulation, resulting in a factor of 2.2 lower mean $\chi^2$ value. One advantage of the new
technique is that it provides the physical parameters on a geometrical height scale. It constitutes a first step towards inversions giving results
consistent with the MHD equations.

\end{abstract}


\section{Introduction}

In recent decades, many insights into photospheric processes were retrieved from spectropolarimetric
observations, mainly in the visible and near-infrared spectral range. Progress in this field was not only reached by the availability of
ever larger telescopes (with an improved spatial resolution) or by going above the Earth's disturbing atmosphere, but also
by advancements in the analysis techniques. While in the early years of spectropolarimetric observations the physical parameters of
the solar atmosphere (temperature, magnetic field vector, line-of-sight (LOS) velocity, pressure, density and their height dependence) were
directly derived from the Stokes profiles \citep[classical estimates, see, e.g.,][for an overview]{Solanki1993}, during the last approximately
25 years the usage of Stokes inversion techniques has become established. Initially the Milne-Eddington approach \citep[involving
height-independent atmospheric parameters; see, e.g,][]{Harvey1972,Auer1977,Borrero2014} was most widely used, but with time (and increasing computing power)
inversion codes that numerically solve the full radiative transfer equations became increasingly important. Prominent examples of such codes
are the SIR code \citep[{\bf S}tokes {\bf I}nversion based on {\bf R}esponse Functions;][]{RuizCobo1992} and the SPINOR code
\citep[{\bf S}tokes-{\bf P}rofiles-{\bf IN}version-{\bf O}-{\bf R}outines;][]{Frutiger2000a,Frutiger2000b}, which assume local thermodynamic
equilibrium (LTE), or the NICOLE code \citep[{\bf N}on-LTE {\bf I}nversion {\bf CO}de using the {\bf L}orien {\bf E}ngine;][]{SocasNavarro2000,SocasNavarro2015},
which also considers non-LTE conditions.

A Stokes inversion is an iterative process that needs a first guess of the atmospheric parameters, which can be obtained from model
calculations, from older results retrieved from similar targets, or in case of the magnetic field vector it can also be simply
constant with height or random. The Unno-Rachkovsky equations for polarized radiative transfer \citep{Unno1956,Rachkovsky1962a,Rachkovsky1962b,Rachkovsky1967}
are numerically solved for the initial atmosphere, which provides the first guess of the synthetic Stokes profiles. These synthetic Stokes profiles
are then compared with observed Stokes profiles. Differences between the two sets of profiles are used to systematically change the initial atmosphere.
This fitting process is iteratively repeated until a good match between the synthetic and observed Stokes profiles is reached. Finally, after multiple
iterative steps, the atmosphere that leads to the best fit is considered as the best representation of the real Sun (within the limitations of the model)
and is output as the result of the Stokes inversion.

Before the observed Stokes profiles can be compared with the synthetic profiles, the latter need to be degraded with all the
instrumental effects that were present during the observation. A convolution of the synthetic Stokes profiles with the spectral transmission
profile of the used instrument (or its approximation by fitting the macroturbulent velocity) is standard in nearly all
inversion codes. Such a transmission profile can be obtained, e.g., by a laboratory measurement as part of the instrument calibration, or by a
comparison of measured spectra with spectrally highly resolved spectra unaffected by spectral stray light \citep[e.g., spectra recorded
with the Fourier Transform Spectrometer, cf.][]{Bianda1998,AllendePrieto2004}.

A further improvement in the quality of the inversion results could be reached by considering the spatial degradation, which introduces a spatial coupling
between the pixels of a data set. Such spatial degradations can be described by the spatial point-spread function (PSF) of the instrument that
can either be measured, e.g., via the phase-diversity (PD) technique \citep{Gonsalves1979,Paxman1992,MartinezPillet2011}, or it can be modeled from
the telescope geometry \citep[e.g.,][]{Danilovic2008}, or determined from an eclipse or a transit of Mercury or Venus
\citep[e.g.,][]{WedemeyerBoehm2008,DeForest2009,Mathew2009}.
The Stokes images can then be deconvolved with the spatial PSF prior to inversions \citep[e.g.,][]{DeForest2009,Beck2011}, which regularly lowers the signal-to-noise ratio
\citep[e.g., in case of the \sunrise{}/IMaX data by a factor of three, see][]{MartinezPillet2011}. In recent years methods were developed to avoid
the increase in noise by considering the spatial PSF not prior to, but during the inversion process.

\begin{figure*}
\centering
\includegraphics[width=\linewidth]{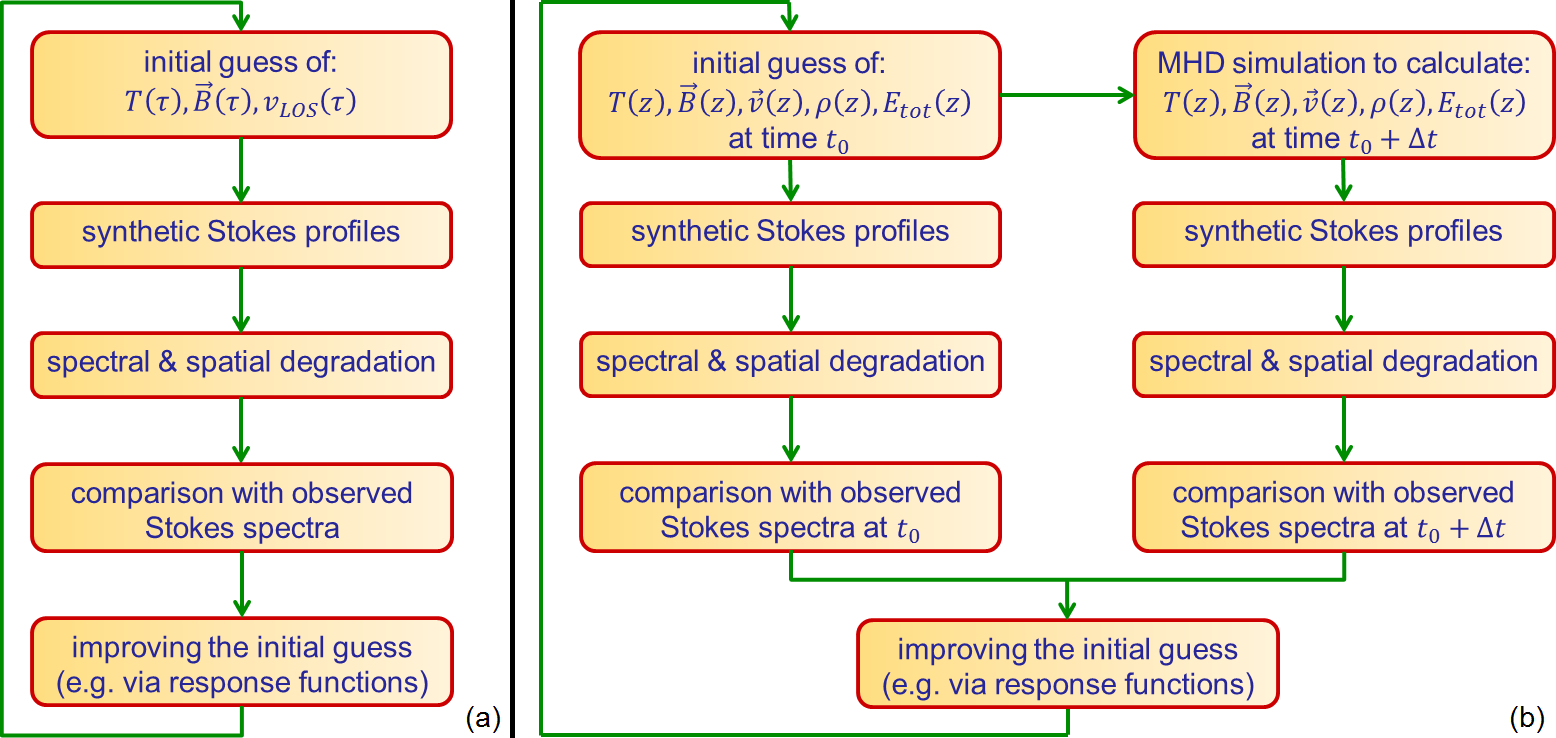}
\caption{Panel a: Flow chart of the iterative fitting process of the spatially coupled Stokes inversion \citep{vanNoort2012}.
Panel b: Flow chart of a hypothetical inversion process that includes the magnetohydrodynamics equations. See the main text for details.}
\label{Fig1}
\end{figure*}

\citet{vanNoort2012} extended the Stokes inversion code
SPINOR by convolving the synthetic Stokes images with a given spatial PSF before the comparison with the observations (see panel a of Fig.~\ref{Fig1}).
Since this approach strictly avoids any deconvolution, the increase of noise can be avoided to a large extent. Data recorded with the spectropolarimeter (SP)
onboard the \textsc{Hinode} satellite \citep{Tsuneta2008,Lites2013} were then successfully used to apply the spatially coupled SPINOR code
\citep[e.g.,][]{Riethmueller2013,Tiwari2013,vanNoort2013,Lagg2014,Buehler2015}. \citet{RuizCobo2013} introduced another approach that decomposes
the Stokes profiles into their principal components, which was later applied to quiet-Sun data recorded with \textsc{Hinode}/SP \citep{QuinteroNoda2015}.
It begs the question what further improvements of the Stokes inversion technique appear possible.

Since the dynamics and evolution of the Sun are of particular interest, often observation data sets do not only consist of individual snapshots
at isolated points in time but of time series. The temporal evolution of the plasma in the solar photosphere can be well described by the equations
of MHD, so that two consecutive observations of a time series (separated by a sufficiently small interval of time, $\Delta t$) are not
statistically independent of each other, but are coupled by the MHD equations (including the equation of energy transfer). We expect that
an inclusion of the MHD equations in the Stokes inversion of a time series will significantly improve the results, because a model that combines
the radiative transfer equations with the MHD equations contains more physics and hence has higher chances of approaching the real Sun.

Panel b of Fig.~\ref{Fig1} demonstrates how the MHD equations can be connected with a Stokes inversion. The process starts with the determination
of an initial guess of the atmospheric parameters for an observation recorded at time $t_0$. In contrast to a traditional Stokes inversion,
the initial guess comprises additional quantities (horizontal velocities, plasma density and total energy) and has to be given on a geometrical
height ($z$) scale, since this is the natural scale of the MHD equations (instead of an optical depth ($\tau$) scale). The extended set of atmospheric
parameters allows the calculation of the temporal evolution of the plasma via an MHD code till $t_0+\Delta t$, which is the observation time of the
next frame of the time series. After a conversion from the geometrical height into optical depth, a subset of the atmospheric parameters is used
to calculate synthetic Stokes profiles, which then can be spectrally and spatially degraded. The left branch of the flow chart in panel~b of
Fig.~\ref{Fig1} illustrates this for the observation recorded at time $t_0$, while the right branch does it for the next frame recorded at time $t_0+\Delta t$. 
A systematic change of the atmospheric parameters tries to minimize the deviation of the synthetic Stokes profiles from the observed ones (for both time steps,
$t_0$ and $t_0+\Delta t$) in an iterative fitting process.

The extension of the atmospheric parameters from five to nine quantities together with the change to a geometrical height scale (of typically
hundreds of grid points instead of just a few optical depth nodes), as well as the need for the relatively slow MHD simulations make the
computational effort of a complete integration of the MHD equations into the Stokes inversion orders of magnitude higher than present
computational capabilities. An implementation of the algorithm as shown in panel b of Fig.~\ref{Fig1} is hence not yet possible. 

This work describes the very first approach towards the full spatially and temporally coupled inversion
of spectropolarimetric observations of the solar photosphere, using MHD simulations and the PSF to link the temporal and spatial information.
Sect.~\ref{Observation} explains the used observations, Sect.~\ref{Simulation} the employed MHD simulations. The method itself is outlined in
Sect.~\ref{Method}, while in Sect.~\ref{Summary} we summarize and discuss the results we have achieved so far.
 
\section{Observations}\label{Observation}

The observations used in this study where recorded in June 2013 during the second stratospheric flight of the balloon-borne solar observatory \sunrise{}.
We refer to \citet{Barthol2011} for technical details of the telescope and the gondola. \citet{Solanki2016} give an overview of the updates applied
to the second flight. Image stabilization, correlation tracking, and real-time sensing of the lower wavefront aberrations were realized by
a Shack-Hartmann wavefront sensor that also controlled the telescope's focus mechanism \citep{Berkefeld2011}.

Two science instruments were operated in parallel onboard \sunrise{}: the \sunrise{} Filter Imager \citep[SuFI;][]{Gandorfer2011} to record
broadband filtergrams in the violet and near-ultraviolet spectral range (214--397)~nm and the Imaging Magnetograph eXperiment \citep[IMaX;][]{MartinezPillet2011}
for spectropolarimetric observations in the Fe\,{\sc i} 525.02~nm line. This spectral line has a Land\'e factor $g=3$ and is one of the most Zeeman-sensitive
lines in the visible spectrum. An overview of the recorded data is given by \citet{Solanki2016}. In this study we refer to an observation recorded by the
IMaX instrument on 2013 June 12, 23:39 UT. The telescope pointed to the active region AR~11768 close to disk center (with cosine of the heliocentric angle $\mu=0.93$).
IMaX was operated in the V8-4 mode, i.e., the full Stokes vector was measured at eight wavelength positions (7 within the spectral line at
$(-12, -8, -4, 0, +4, +8, +12)$~pm and one in the continuum at $+22.7$~pm offset from the line core) where four images of 250~ms exposure time were
accumulated at a time. The plate scale of IMaX was 0\carcsec{}0545 per pixel.

In the post-processing of the IMaX data, corrections for dark current and flat-field effects where applied. A pre-flight polarimetric calibration
provided a M\"uller matrix for each pixel of the field of view (FOV) that allowed a removal of the instrumental polarization \citep{MartinezPillet2011}.
Because the primary mirror could not be included in the polarimetric calibrations and because the thermal environment was different during the observations
than during calibration, the elements of the M\"uller matrices slightly changed, which we took into account by applying an ad-hoc cross-talk removal
of the order of 1\,\%. The thus reduced data are named level-1 (or non-reconstructed data).

A PD measurement was achieved about two hours before the observation by inserting a PD plate into the optical path of one of the two
IMaX cameras. This allowed the retrieval of a PSF used to correct the data for low-order wavefront aberrations \citep{Gonsalves1979,Paxman1992}.
The PD reconstructed data are named level-2. More details are reported by \citet{MartinezPillet2011}. See also \citet{Solanki2016} for additional details
concerning data reduction steps specific to IMaX during the second \sunrise{} flight.

\section{Simulation and spectral synthesis}\label{Simulation}

We accomplished realistic simulations of the radiative and magneto-hydrodynamical processes in the solar photosphere and upper convection zone with the
MURaM ({\bf M}ax Planck Institute for Solar System Research / {\bf U}niversity of Chicago {\bf Ra}diation {\bf M}agneto-hydrodynamics) code,
a three-dimensional, non-ideal, compressible MHD code that includes non-gray radiative transfer calculations in the energy equation under the assumption of
local thermal equilibrium \citep{Voegler2005}. The simulation box is 24~Mm~$\times$~24~Mm in its horizontal dimensions and has a depth of 6.1~Mm. The cell
size of the simulation box is 41.7~km in the two horizontal directions and 16~km in the vertical direction.

We used a statistically relaxed purely hydrodynamical simulation as initial condition. A homogeneous unipolar vertical magnetic field of $\langle B_{\rm{z}} \rangle = 400$~G
was then introduced into the hydrodynamical simulation and the simulation run was continued for another three hours of solar time to reach again a statistically relaxed
state. After the relaxation a single snapshot covering several pores with field strengths up to 3500\,G (at $\log(\tau)=-1$) is used for this study,
see the left panel of Fig.~\ref{Fig2}. The boundary conditions where periodic in the horizontal directions, closed at the top boundary of the box, while a free in- and
outflow of plasma was allowed at the bottom boundary under the constraint of total mass conservation. The $\tau=1$ surface for the continuum at 500~nm was on average
reached about 700~km below the upper boundary.

The forward calculation mode of the SPINOR inversion code \citep{Solanki1987,Frutiger2000a,Frutiger2000b} was used to compute synthetic Stokes spectra of the
Fe\,{\sc i} 525.02~nm line used by IMaX. We synthesized the strongest 20 spectral lines in the range (524.72--525.32)~nm whose atomic parameters are
listed in \citet{Riethmueller2014}. The radiative transfer was calculated for a heliocentric angle of $\mu=0.93$.

\section{Method}\label{Method}

\subsection{Concept}\label{Concept}

In this section we present an alternative method to obtain the atmospheric parameters from observed Stokes profiles, i.e., an alternative to the traditional
Stokes inversion technique purely based on radiative transfer calculations and the computation of a model atmosphere assuming hydrostatic equilibrium
(usually restricted to the vertical direction). The idea is to use an MHD simulation of a target similar to the observations (in our case pores and granulation)
for a Stokes synthesis of the observed spectral range. The synthetic Stokes data then have to be adapted to the pixel size of the observation and degraded
with all the instrumental effects that influenced the observational data. To gain a comprehensive knowledge about these instrumental effects is possibly
the largest difficulty of the method.

We now consider the degraded MHD data set as a pool of Stokes profiles that can be directly compared with the observed Stokes profiles. In a first iteration,
a pixel in the degraded data set is assigned to a particular observed pixel, namely the one that shows the least deviation between observed and synthetic
Stokes profiles. The assigned MHD pixels embody the inversion result of the first iteration.

Our inversion result (a pixel in the degraded MHD data set) is not only connected to the temperature, magnetic field vector,
and LOS velocity in the photosphere as is the case for results of traditional Stokes inversion techniques, but also the full set of undegraded MHD data is available,
including horizontal velocities and also including information on the layers immediately below the solar surface, which are not directly accessible via
spectral lines. Because the layers below the solar surface drive the features at the solar surface, the inversion of a time
series of Stokes data has the potential to constrain the sub-surface dynamics. The stratifications of the atmospheric parameters obtained by our inversion method are available on a geometrical height scale, while
traditional Stokes inversions usually provide an optical depth scale, whose conversion into geometrical heights is influenced by the underlying
assumptions \citep[see, e.g.,][]{Puschmann2010}.

While the method up to this point has already been considered by other authors in a more or less similar way
\citep{MolownyHoras1999,Tziotziou2001,Berlicki2005,Carroll2008,Beck2013b,Beck2015}, we go a step further and re-sort the MHD data set according to
the best-fit results (see Sect.~\ref{MASI}) and use the re-sorted data as initial condition of a new MURaM simulation (see Sect.~\ref{MuramContinuation}).
The re-sorting takes a physically consistent simulation box and produces a physically inconsistent one out of it, e.g. the horizontal flow pattern and
the magnetic field lines are destroyed by the rearranging. During a relatively short relaxation process the MURaM code removes the physical inconsistencies
(in the same way when the 400\,G were artificially introduced, see Sec.~\ref{Simulation}) and we obtain MHD data that are very similar to the observation.

While we only required from the original MHD simulation that it contains granulation and pores of any size and shape, the new simulation
shows pores having sizes, shapes and positions within the FOV that are similar to the observation. The better an MHD simulation matches
the observation the better the fitting of Stokes profiles can work, so that a significant improvement of the results can be reached by
repeating the method iteratively, i.e. the new MHD simulation is used as input for a second iteration of spectral synthesis, degradation,
Stokes profile fitting, and MHD simulation.

We name the new technique MHD-Assisted Stokes Inversion (MASI).

\subsection{Degradation of the MHD data}\label{Degradation}

As mentioned in Sect.~\ref{Observation}, we have the choice between non-reconstructed and reconstructed observations.
The reconstruction is done with a spatial PSF determined by in-flight PD measurements which provide the low-order aberrations of the telescope
(defocus, coma, astigmatism, etc.), i.e., aberrations that determine the inner core of the PSF. The PD measurements do not (or hardly) provide information
on high-order aberrations which determine the wings of the PSF, e.g., the stray-light behavior of the system.

Only if the degradation of the synthetic data reflects the real situation in the observed data to a high degree a physically meaningful comparison between them
is possible. From the higher quality of the reconstructed data it is clear that a comparison with the non-reconstructed data requires a stronger degradation
than a comparison with reconstructed data. Because deconvolution of the data with the PD PSF is not a straightforward procedure, we demonstrate our new
inversion technique with non-reconstructed data. This is to strictly avoid any deconvolution of the observations. In the following we depict all
steps needed to degrade the synthetic data to the level of the non-reconstructed observations (level-1).

\subsubsection{Spectral resolution and sampling}

The spectral PSF of IMaX was measured in the laboratory before the launch of \sunrise{} \citep[see the bottom panel of Fig.~1 in][]{Riethmueller2014}
and considers not only the transmission behavior of the etalon but also the used pre-filter. In addition to the 6.5~pm wide primary peak, 
the spectral PSF also shows secondary peaks at around $\pm 200$~pm offset from the position of the transmision maximum. In order to include a possible
influence of the secondary peaks we synthesize all the 20 spectral lines we could identify in the $\pm 300$~pm range around 525.02~nm.

The synthetic Stokes profiles are convolved with the spectral PSF in order to reproduce the spectral resolution of IMaX. Finally, Stokes images are
created for the spectral range $-15$~pm to $+30$~pm in spectral steps of $1$~pm, because such two-dimensional Stokes images are needed for a
convolution with a spatial PSF such as the one we apply in the following degradation step. The chosen spectral sampling of $1$~pm is a reasonable compromise
between the amount of data and the spectral resolution. The chosen spectral range is a bit wider than the scanning range of IMaX in its V8-4 mode
($-12$~pm to $+22.7$~pm) because of the blueshift effect inherent in IMaX, which will be explained in section~\ref{MASI}.

\subsubsection{Diffraction limit and low-order aberrations}

As mentioned above, the diffraction at the telescope 1-m aperture and the low order wavefront aberrations were measured by IMaX during
the flight via the phase-diversity technique. Consequently, the synthetic Stokes images are convolved with the spatial PSF retrieved from
the PD data to take these effects into account.

\subsubsection{Residual jitter}

Besides the 1-m aperture of the telescope at the considered wavelength, the spatial resolution of \sunrise{} was limited
by the stability of the image. The pointing system stabilized the gondola to an accuracy of a few arcseconds,
while the Correlating Wave-Front Sensor (CWS) controlled a tip/tilt mirror so that the reflected beam was further
stabilized down to an RMS value of only 0\carcsec{}025 (measured during the flight by the CWS). Applying an
artificial jitter of that strength to MHD images led to significantly less image smearing than seen in the IMaX observations.
A possible, but speculative, explanation could be a differential motion between the CWS camera and the IMaX cameras. All cameras are
located within the post-focus-instrumentation (PFI) structure, that is made of carbon fiber and was designed for high
stiffness. Nonetheless, with an F/121 beam inside the Image Stabilization and Light Distribution System \citep[ISLiD;][]{Gandorfer2011}
even a minute bending (due to vibrations) would suffice to explain the IMaX image quality.

In order to take the actual image smearing into account we decided to apply a combination of two Butterworth lowpass filters
\citep[one with a low cut-off frequency, the other with a much higher one;][]{Butterworth1930} because it degraded the MHD data slightly
more realistically than a convolution with a two-dimensional Gaussian \citep[as we applied to data from the first \sunrise{} flight in][]{Riethmueller2014},
in particular with respect to the power spectrum.

\subsubsection{Stray light}\label{StrayLight}
A reliable determination of the stray-light properties of \sunrise{} is a challenge since they strongly
influence the data (in particular the RMS intensity contrast) but are difficult to measure. \citet{Riethmueller2014}
assessed the stray-light contamination of the first flight IMaX data from analyzing observations
of the solar limb. Unfortunately, during the second flight the limb was only observed in the late phase of the
mission when IMaX had problems with the electronics stabilizing the etalon temperature.

The limb data contain clear evidence for a non-local stray-light contamination, i.e., that the stray-light PSF has extended
wings. Additionally, we can distinguish between two stray-light sources: The light level seen at the image border not exposed
to sunlight (the border is caused by the field stop at the entrance of the IMaX instrument) can only be caused by IMaX internal stray light,
while the somewhat higher total stray-light level in the off-limb region of the images indicates a second, albeit smaller contribution coming
most probably from outside IMaX. Unfortunately, we cannot use the limb data for a more quantitative assessment of the stray-light properties
because of the floating etalon temperature and also because the correlation tracker had problems to stabilize the
image in the direction parallel to the solar limb, making these images less sharp than those on disk.

During the \sunrise{} calibration on ground, a fibre bundle was placed in the secondary focal plane, F2. The fibre bundle
is an extended light source with a diameter of about 8\carcsec{}5 centered in the $51\arcsec{} \times 51\arcsec{}$ IMaX FOV
making this a nearly ideal target to determine the stray-light properties of IMaX. An illumination with sunlight
was not possible for technical reasons so that an artifical light source (75~W Xenon lamp) had to be used. All frames
of the fibre-bundle calibration campaign were summed up. The summed image was then deconvolved with the ideal (i.e. stray-light
and aberration free) fibre-bundle image approximated by a binary mask image retrieved by applying an intensity threshold to
each fibre of the summed fibre-bundle image. The deconvolution was done via the Lucy-Richardson algorithm \citep{Richardson1972,Lucy1974}
which is an iterative method that uses only convolutions. Compared to the Sun, the light level of the used artifical light source
was extremely low leading to a low signal-to-noise ratio in the outer parts of the fibre-bundle images. Hence the stray-light related
far wings of the PSF could not be retrieved with the needed accuracy.

Even if these attempts in measuring the stray-light PSF did not yield a result that could be directly used, they helped to
considerably narrow down the shape of the azimuthally averaged PSF. From this we could determine that the stray light seen by IMaX has
neither a purely local nor a purely global character, but that an influence of the stray light on every spatial scale is present.

Since the stray-light PSF could not be determined with adequate accuracy, only the utilization of a simplistic stray-light model remains.
We contaminate the synthetic Stokes~$I$ images with 25\,\% global stray light.\footnote{Contamination with 25\,\% global stray light means that a
subtraction of 25\,\% of the spatial mean Stokes profile from the individual contaminated profiles results in the original (stray-light free) profiles.}
Stokes~$Q$, $U$, $V$ are not contaminated because most of the stray light was created internally inside IMaX and the internal stray light is
largely unpolarized.

The above strength of the global stray light was tested by a test-wise inversion of the IMaX data carried out with the traditional SPINOR code and in
particular by evaluating the resulting magnetic field maps. A simple and hence robust atmospheric model was used: Three $\log(\tau)$ nodes for the
temperature; everything else was assumed to be height-independent. Without any stray-light correction the inversion provided kilogauss fields mainly
along the edges of pores, and unrealistic weaker fields in the center of the pores. This effect can be explained by the broad wings of the stray-light PSF.
Inside the pores the low temperatures together with the shallow temperature gradients lead to low intensities and extremely small line depths of the 525.02~nm
spectral line, so that a contamination with light from the bright vicinity of the pores causes a significant underestimation of the field strength.
A global stray-light correction of the data lowered this effect considerably. The best results could be reached for a stray-light strength of 25\,\%.

Two other methods, which are independent of the field weakening in the pores, confirmed that a 25\,\% of global stray-light together with the
Butterworth filtering is reasonable. Firstly, the intensity of the darkest pixel in the synthetic pores in the MHD simulations increases
from 8.6\,\% to 38.8\,\% of the mean quiet-Sun intensity, $I_{\rm{QS}}$ when the various degradation steps described above are applied. This value
agrees nicely with the minimum intensity of the non-reconstructed IMaX data, 38.4\,\%. We note that the size of the largest pore in the MHD data
is smaller than in the IMaX data, which possibly makes the comparison of the minimum intensities problematic, but later in this study we will see
that the match also holds if the IMaX data are compared with simulations of nearly identical pore sizes. Secondly, the RMS contrast of the Stokes~$I$
continuum image of the non-reconstructed IMaX data, 6.8\,\%, is similar to the constrast of the fully degraded MHD data, 7.7\,\% (pores were excluded
when we determined these numbers).

\subsubsection{Noise}
IMaX Stokes~$Q$, $U$, and $V$ images at the continuum wavelength (+22.7~pm) were used to calculate histograms of a quiet-Sun region,
which is generally free of polarization signals. The histograms showed a clear Gaussian shape with a standard deviation of about
$2.3 \times 10^{-3} I_{\rm{QS}}$ (level-1 data) or $7 \times 10^{-3} I_{\rm{QS}}$ (level-2 data), respectively. $I_{\rm{QS}}$ is the mean
continuum intensity of the selected quiet-Sun region. We did not add any noise to the synthetic data to avoid complicating the comparison with
the observations, but we used the standard deviation as the noise value for all four Stokes parameters when calculating the
$\chi^2$ values of the merit function we define in Eqs.~(\ref{Eq_Merit1}) and (\ref{Eq_Merit2}).

\subsection{Assessment of the degradation}

Since our method relies on a direct comparison between degraded synthetic Stokes profiles and observed Stokes profiles
\citep[see also][]{Danilovic2010,Beck2013a,Riethmueller2014}, a realistic degradation is crucial.
In this section we assess the quality of the degradation steps decribed in Sect.~\ref{Degradation} by comparing the two data sets. In our experience it is
important that not just a single quantity (e.g. the RMS intensity contrast) is compared but multiple ones. In practice, the measurement of instrumental
effects might not be possible with the necessary accuracy, so those effects often need to be approximated by simplified models (in our case stray light and jitter)
containing one or more adjustable parameters.

\begin{figure*}
\centering
\includegraphics[width=\linewidth]{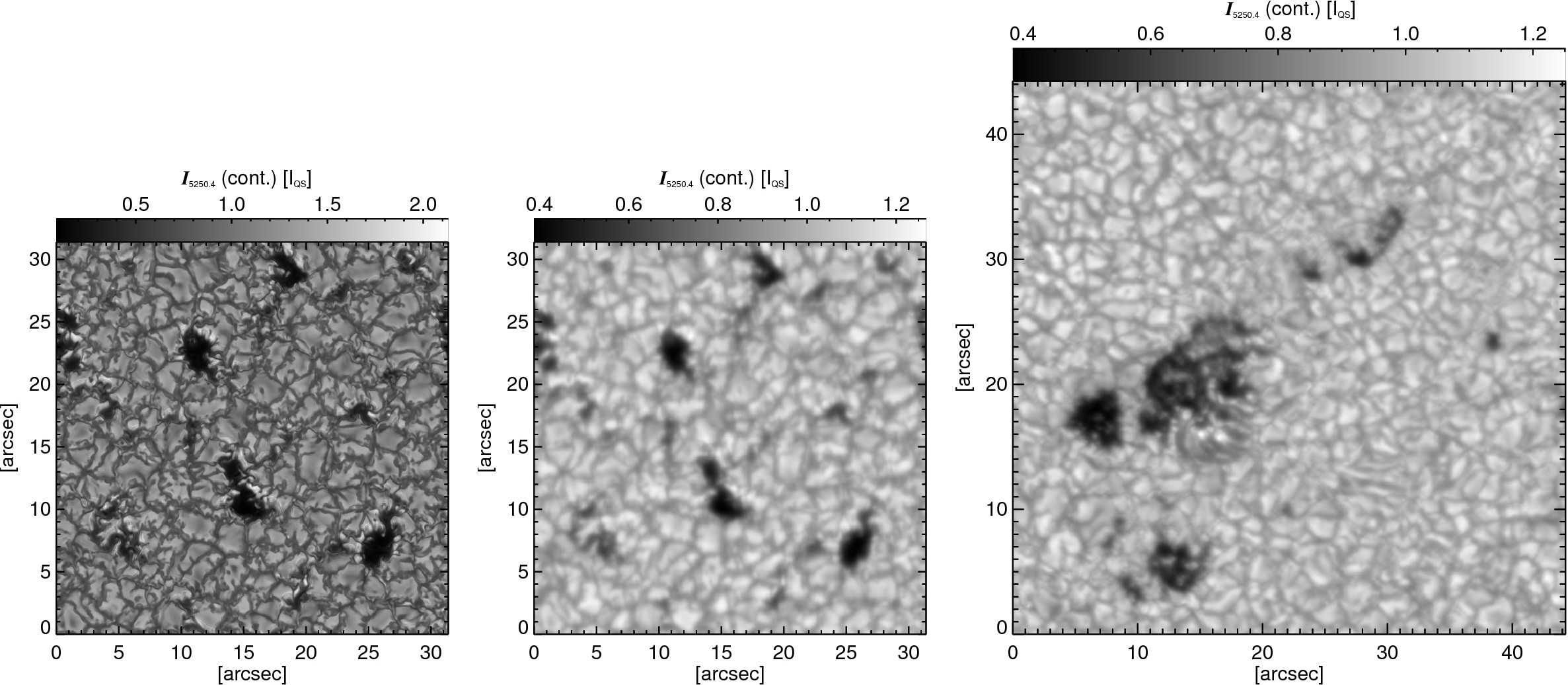}
\caption{Images of the Stokes~$I$ continuum at 525.04~nm, normalized to the mean quiet-Sun intensity, $I_{\rm{QS}}$. The left panel displays
the original (i.e. undegraded) MHD data, the middle panel the fully degraded MHD data, and the right panel the level-1 IMaX observation.}
\label{Fig2}
\end{figure*}

Fig.~\ref{Fig2} exhibits the influence of the degradation on the synthetic Stokes~$I$ continuum image and compares the undegraded and degraded synthetic images
with the corresponding IMaX observation. The degradation clearly lowers the contrast and the spatial resolution so that the image quality of the synthetic data
is brought down to the level of the observation.

\begin{figure}
\centering
\includegraphics[width=\linewidth]{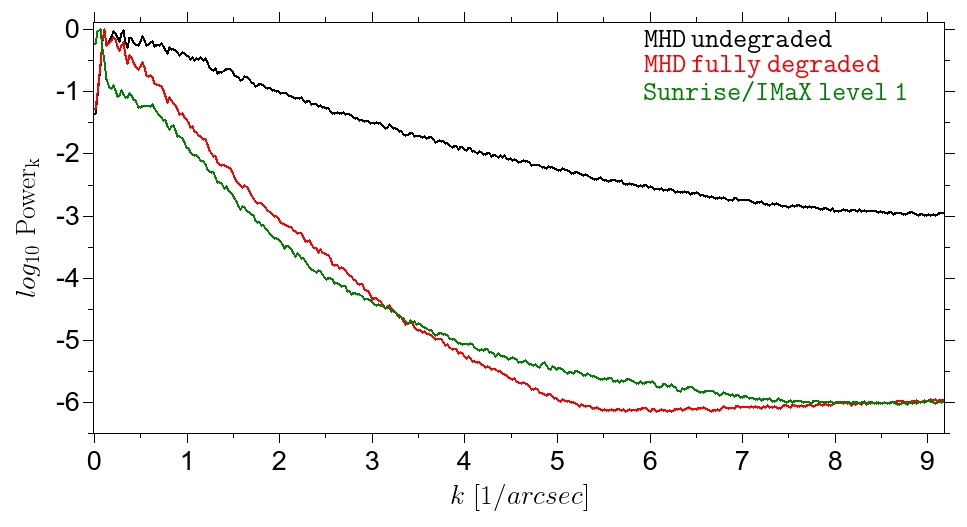}
\caption{Logarithm of the azimuthally averaged power spectra of the Stokes~$I$ continuum images versus the wave number, $k$.
The black line corresponds to the undegraded MHD simulation, the red line to the fully degraded data, and the green line displays
the level-1 IMaX observation.}
\label{Fig3}
\end{figure}

\begin{figure}
\centering
\includegraphics[width=\linewidth]{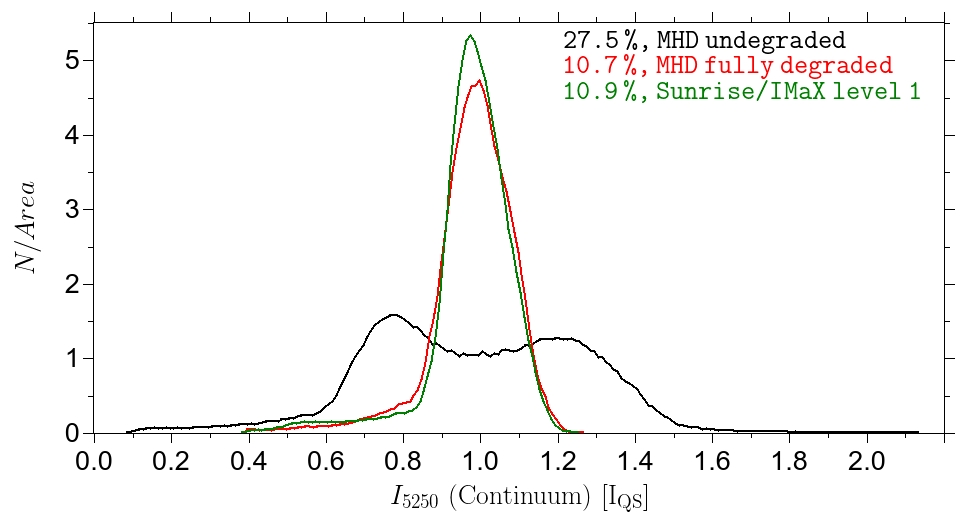}
\caption{Histograms of the normalized Stokes~$I$ continuum intensity. The color coding is the same as in Fig.~\ref{Fig3}. RMS contrasts are indicated in the text labels.
The histograms cover the full FOV, i.e. the pores are included.}
\label{Fig4}
\end{figure}

\begin{figure}
\centering
\includegraphics[width=\linewidth]{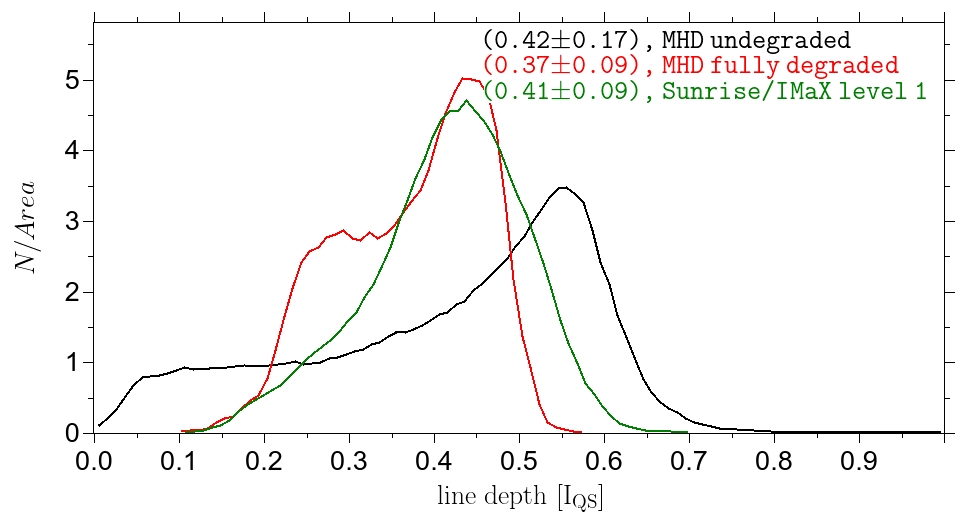}
\caption{Same as Fig.~\ref{Fig4}, but for the Stokes~$I$ line depth. Mean values and their standard deviations are indicated in the text labels.}
\label{Fig5}
\end{figure}

In Fig.~\ref{Fig3} we show the azimuthally averaged power spectra \citep[e.g.,][]{Puschmann2011,Katsukawa2012} and in Fig.~\ref{Fig4} the
intensity histograms both calculated from the images displayed in Fig.~\ref{Fig2}. The same color coding is used for Figs.~\ref{Fig3}~to~\ref{Fig5}:
The black line corresponds to the undegraded synthetic data, the red line to the degraded synthetic data, and the green line to the non-reconstructed
IMaX observations.

A comparison of the power spectra reveals that our degradation steps bring the spectrum down to roughly the level of the observed spectrum,
but some mismatch remains. For wave numbers $k<3.5~\mathrm{arcsec^{-1}}$ the synthetic power is not sufficiently suppressed by the degradation,
while it is too low for higher wave numbers (see Fig.~\ref{Fig3}).

The degradation of the synthetic data reduces the RMS intensity contrast from 27.5\,\% down to 10.7\,\% which is 0.2\,\% lower than the observational
contrast of 10.9\,\% (see Fig.~\ref{Fig4}). The histograms are calculated for the full FOV, including pores. If we exclude the pores
(that fill different fractions of the FOV in the two data sets) then we find the RMS contrast values given in Sect.~\ref{StrayLight}.

As mentioned above, the stray light can have a severe influence on the line depths, in particular in the pores. Hence we also show
histograms of the Stokes~$I$ line depth in Fig.~\ref{Fig5}. While the standard deviation matches well between the degraded simulation and the observation,
the mean value of the synthetic data is slightly lower than the observed value (see text labels). Additionally, the synthetic line-depth histogram
exhibits a bi-modal shape which is not the case for the observed histogram.

Figs.~\ref{Fig2}~to~\ref{Fig5} illustrate the quality of our degradation. The remaining discrepancies can be caused by various effects, such as
the uncertainties in the stray-light correction mentioned in Sect.~\ref{StrayLight}. Additionally, the lowpass filtering with a combination
of two Butterworth filters is only a very simple way of modelling the residual jitter. Although the degradation of the MHD data might not perfectly
mimic the behavior of IMaX we believe it successfully accounts for the most important contributions.

\subsection{Stokes profile fitting}\label{MASI}

After taking the various instrumental effects into account, we now have two data sets available that are directly comparable, except for two remaining discrepancies.
The first discrepancy concerns the magnetic polarity, because our MHD simulation is a unipolar one, while both magnetic polarities are present in the IMaX observation.
Hence we create a second MHD data set by flipping the sign of the magnetic field vector, while all other quantities remain unchanged. The second MHD data set
is degraded in the same manner as the original one, so that a bipolar data set can be produced by combining the two opposite polarity data sets.

The second discrepancy concerns the blueshift that is the sum of the field-dependent etalon blueshift and the constant convective blueshift.
The etalon blueshift is a wavelength shift across the FOV due to the inclination of the off-axis rays as they reach the etalon in a collimated configuration.
The map of the total blueshift is provided by the IMaX calibration process \citep{MartinezPillet2011}.

For the traditional Stokes inversion method the velocity map needs to be corrected for the blueshift just at the very end. In the MASI case
the blueshift has to be taken into account directly during the fitting of the observed Stokes profiles.

Consider now a given pixel of the observation data. Its position within the FOV of IMaX determines its blueshift. This particular blueshift
is subtracted from the nominal wavelengths of the V8-4 mode. Every degraded synthetic profile (which has high spectral sampling) of all pixels
in the MHD FOV is then re-sampled at the corrected wavelengths, resulting in a blueshift-corrected MHD data set. We then browse through the
corrected MHD data set in order to identify the pixel whose Stokes profiles match the observed one best. We assess the similarity of Stokes profiles
by defining the following $\chi^2$ merit function:
   \begin{equation}\label{Eq_Merit1}
   \chi^2(x_o,y_o)=\frac{1}{32} \sum_{w=1}^{8}{ \sum_{s=1}^{4}{\frac{(I_{w,s}^{\rm{obs}}(x_o,y_o)-I_{w,s}^{\rm{syn}}(x_m,y_m))^2}{\sigma_s^2}}}~,
   \end{equation}
where $(x_o,y_o)$ is the pixel position within the observed FOV composed of $812 \times 812$ pixels, while $(x_m,y_m)$ is the position
within the MHD FOV consisting of $1152 \times 576$ pixels. $w$ runs over the eight
wavelengths of the V8-4 mode and $s$ over the four Stokes parameters, $I$, $Q$, $U$, $V$, e.g. $I_{2,1}^{\rm{syn}}$ means the degraded synthetic
Stokes~$Q$ signal at the wavelength $-4$~pm. $\sigma_s$ is the noise level of the Stokes parameter $s$.

This step of the MASI algorithm searches for the MHD pixel whose degraded Stokes profiles provide the smallest $\chi^2$ value for the given
observed pixel. The position of the best-fit MHD pixel within the MHD FOV is assigned to the observed pixel. Repeating this procedure for all
observed pixels needed 19 hours on a single Intel(R) Core(TM) i7-2760QM and results in two maps having the size of the observed
FOV and giving the x and y coordinates of the best-fit MHD pixels (not shown). These coordinates are then used to create a new MHD data set by
re-sorting the original one. The new MHD data set covers the FOV of the observation and holds for each position within the FOV the one MHD pixel
that fits the observed one at this position best. We note that not all pixels of the simulation need to find their way into the re-sorted data set,
whereas others may provide the best fit to multiple Stokes profiles.

\begin{figure*}
\centering
\includegraphics[width=\linewidth]{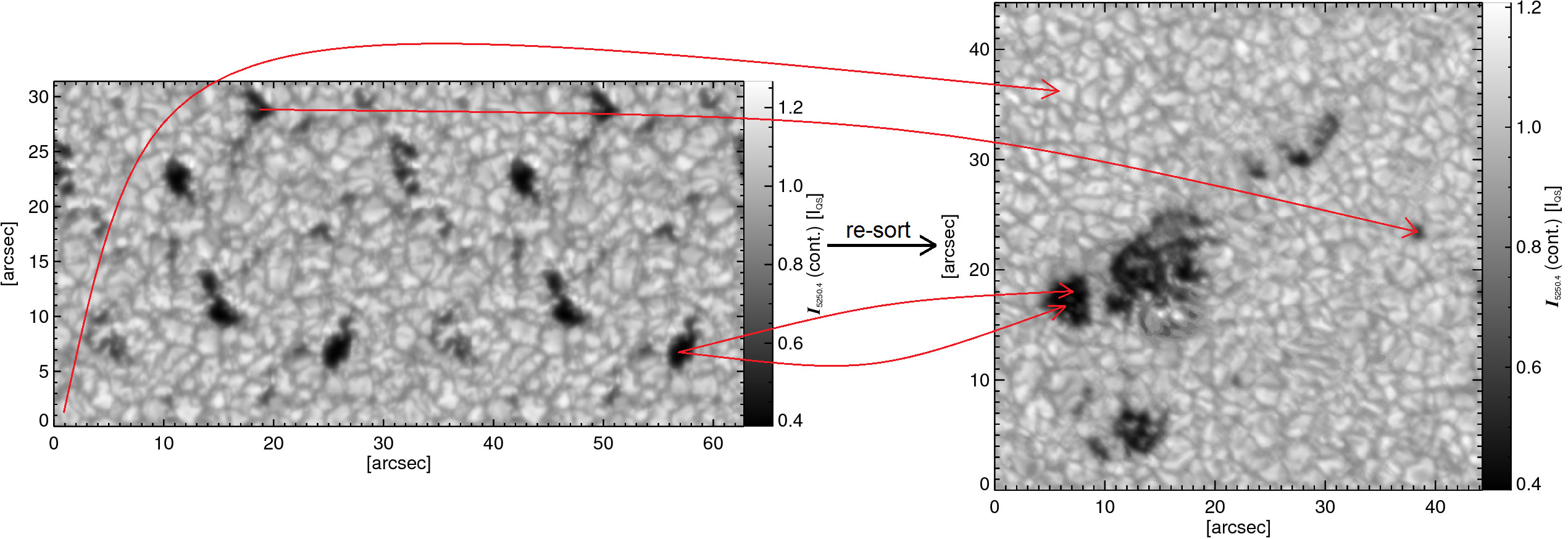}
\caption{Stokes~$I$ continuum image of the degraded MHD data before the re-sorting by the first iteration of the MHD-Assisted Stokes Inversion (MASI) method
(left panel) and afterwards (right panel). The red arrows show a few examples of the re-sorting.}
\label{Fig6}
\end{figure*}

Fig.~\ref{Fig6} demonstrates the re-sorting using the example of the degraded Stokes~$I$ continuum image. The archive of
the Stokes~$I$, $Q$, $U$, and $V$ profiles calculated from the used MHD snapshot contains $2 \times 576 \times 576$ entries (see the left
panel of Fig.~\ref{Fig6}). 19\,\% of the entries in the archive provide best fits to the Stokes parameters in the
IMaX data set, i.e. this fraction is actually chosen by the code to build up the re-sorted data set (right panel
of Fig.~\ref{Fig6}), whose comparison with the IMaX observation (right panel of Fig.~\ref{Fig2}) shows a remarkable match. Only the
penumbra-like region around the position (15\arcsec,14\arcsec) looks less smooth than other regions, mainly because it is
composed of just a few intensity levels.

\begin{figure}
\centering
\includegraphics[width=\linewidth]{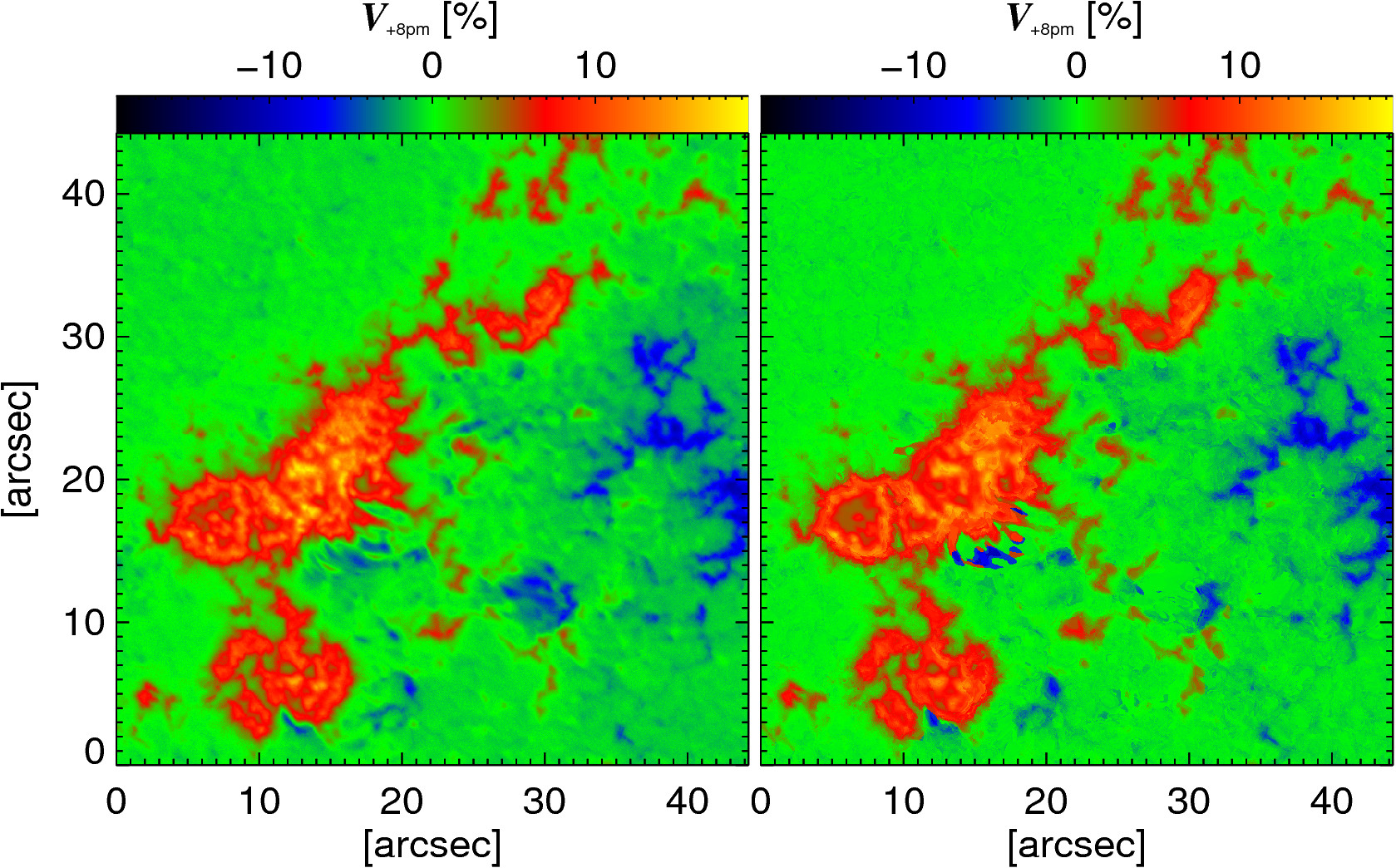}
\caption{Stokes~$V$ image in the red wing of the Fe\,{\sc i} 525.02\,nm line, normalized to the mean quiet-Sun intensity. Left panel: Non-reconstructed data
observed by IMaX. Right panel: Degraded MHD data re-sorted by the first MASI iteration.}
\label{Fig7}
\end{figure}

Maps of the Stokes~$V$ signal at a wavelength offset by $+8$~pm from the line core are compared between the IMaX observation and the MASI result in Fig.~\ref{Fig7}.
The bipolar structure of AR~11768 is well reproduced by MASI as well as most of the magnetic fine structure. A mismatch is again found in the penumbra-like region
around the position (15\arcsec,14\arcsec) and also in the flux-emergence region around (30\arcsec,12\arcsec) because our archive of synthetic
Stokes profiles does not contain profiles of such features with a nearly horizontal magnetic field.

Another difference in the Stokes~$V$ images can be found for the interior of the pores. The peripheral region of the pores is well reproduced by
the first MASI iteration, while the weakening of the circular polarization in the central part of the pores differs slightly. We speculate that the
discrepancy is caused by imperfections in the degradation of the data, in particular by our imperfect knowledge of the stray-light
contribution to the PSF. On top of that, the equal noise levels in the calculation of the merit function leads to a certain bias towards Stokes~$I$
because the Stokes~$I$ amplitude is larger then the Stokes $Q$, $U$, and $V$ amplitudes.

\begin{figure}
\centering
\includegraphics[width=\linewidth]{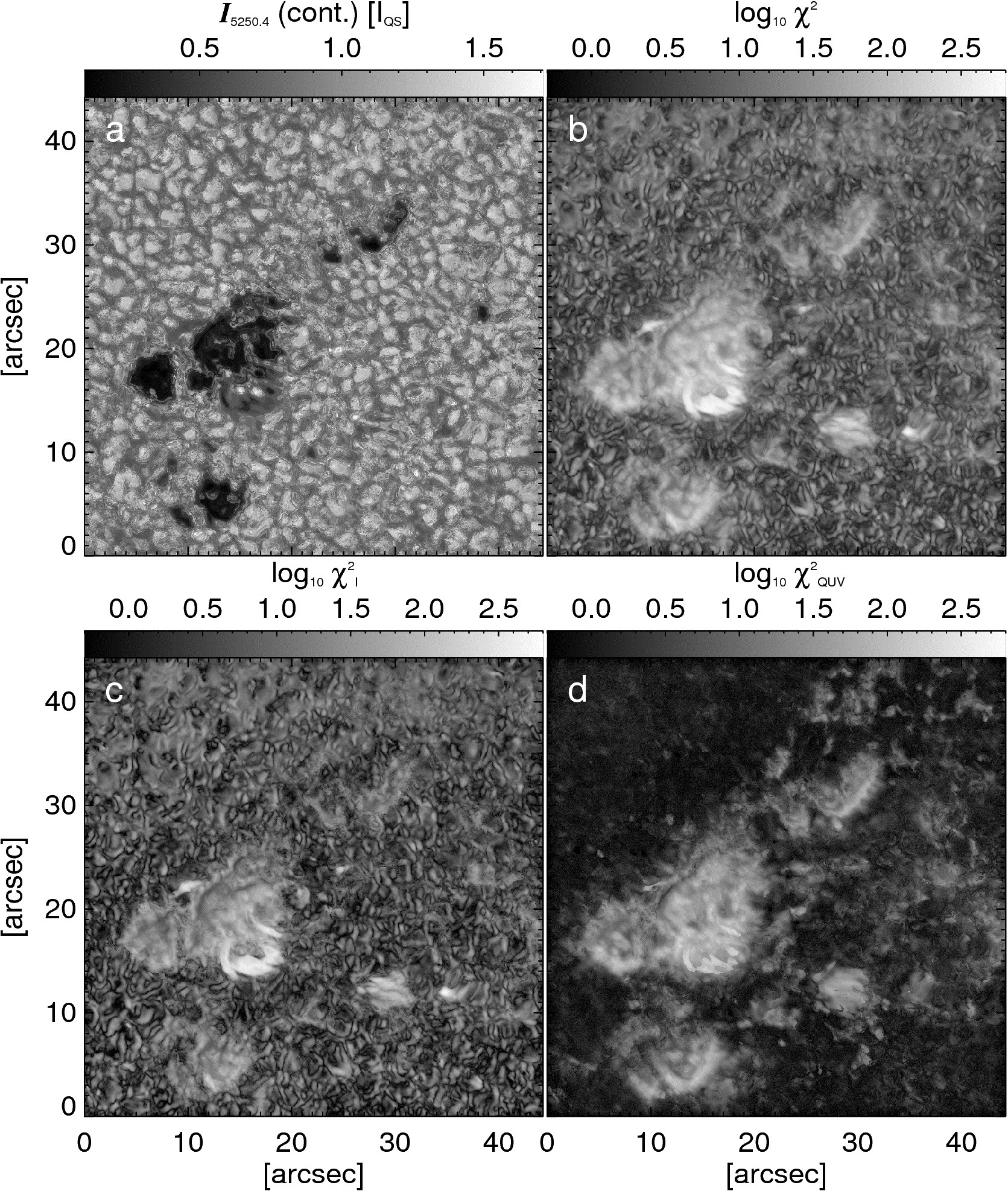}
\caption{Stokes I continuum image taken from the undegraded MHD data set after the re-sorting by the first MASI iteration
(panel~a), map of the total $\chi^2$ values at the end of the first iteration of Stokes profile fitting (panel~b), contribution of Stokes~$I$
(panel~c) and of Stokes~$Q$, $U$, $V$ together (panel~d) to the total $\chi^2$. Panels b-d are plotted on a logarithmic scale.}
\label{Fig8}
\end{figure}

Panel a of Fig.~\ref{Fig8} shows the undegraded Stokes~$I$ continuum image after the re-sorting. Its RMS quiet-Sun intensity contrast
of 22.5\,\% is only slightly below the value of the undegraded MHD data before the re-sorting, 25.1\,\% (left panel of Fig.~\ref{Fig2}) but
considerably larger than the contrast of the degraded image, 6.6\,\% (right panel of Fig.~\ref{Fig6}). The undegraded MASI result restores the
rough topology of pores and granules quite well. However, the transition regions between granules and intergranular lanes are not very smooth
but show lots of small-scale spatial discontinuities.

The quality of the MASI results relies not only on how well the MHD data are degraded but also on a good representation of the observed target types
by the MHD data set. Pores and granulation are contained in our MHD simulation and hence these features can be reproduced by the MHD pixels quite well. Since
the simulation does not contain any penumbra or similar features of more horizontal fields, one cannot expect good fits for the observed penumbra-like region.
A look at the $\chi^2$ map (panel b of Fig.~\ref{Fig8}) reveals the same conclusion. The largest $\chi^2$ values can be found in the penumbral region. The
mean $\chi^2$ value of the first MASI iteration is 17.1, while it is 27 for a traditional SPINOR inversion of the used non-reconstructed IMaX observation (including
a 25\,\% global stray-light correction). A decomposition of the $\chi^2$ values into contributions from Stokes~$I$ (panel c of Fig.~\ref{Fig8})
and Stokes~$Q$, $U$, and $V$ together (panel d of Fig.~\ref{Fig8}) reveals that even for the regions of more horizontal field around positions (15\arcsec,14\arcsec)
and (30\arcsec,12\arcsec) the two contributions are of roughly the same magnitude. On average Stokes~$I$ contributes 63\,\% to the total $\chi^2$ value,
while the fraction of Stokes~$Q$, $U$, and $V$ together is 37\,\%.

More details of the re-sorted MHD data will be presented after the second MASI iteration in Sect.~\ref{Iterating}.   

\subsection{Continuation of the MHD simulation}\label{MuramContinuation}

After the first step of the MASI method has been introduced and applied to an IMaX observation and after we have presented some results,
we now take the next step, which takes advantage of the availability of the entire set of MHD quantities because the re-sorting does not
only process the degraded Stokes images but all available data are re-sorted. We start a new MURaM simulation
that uses the first step MASI result as the initial condition. The boundary conditions of the new run are identical to the ones
described in Sect.~\ref{Simulation}. The re-sorting of the MHD data increased the horizontal dimensions of the simulation box to
the size of the IMaX observation, 33.8~Mm~$\times$~33.8~Mm, while the depth of the box is kept at 6.1~Mm.

\begin{figure*}
\centering
\includegraphics[width=\linewidth]{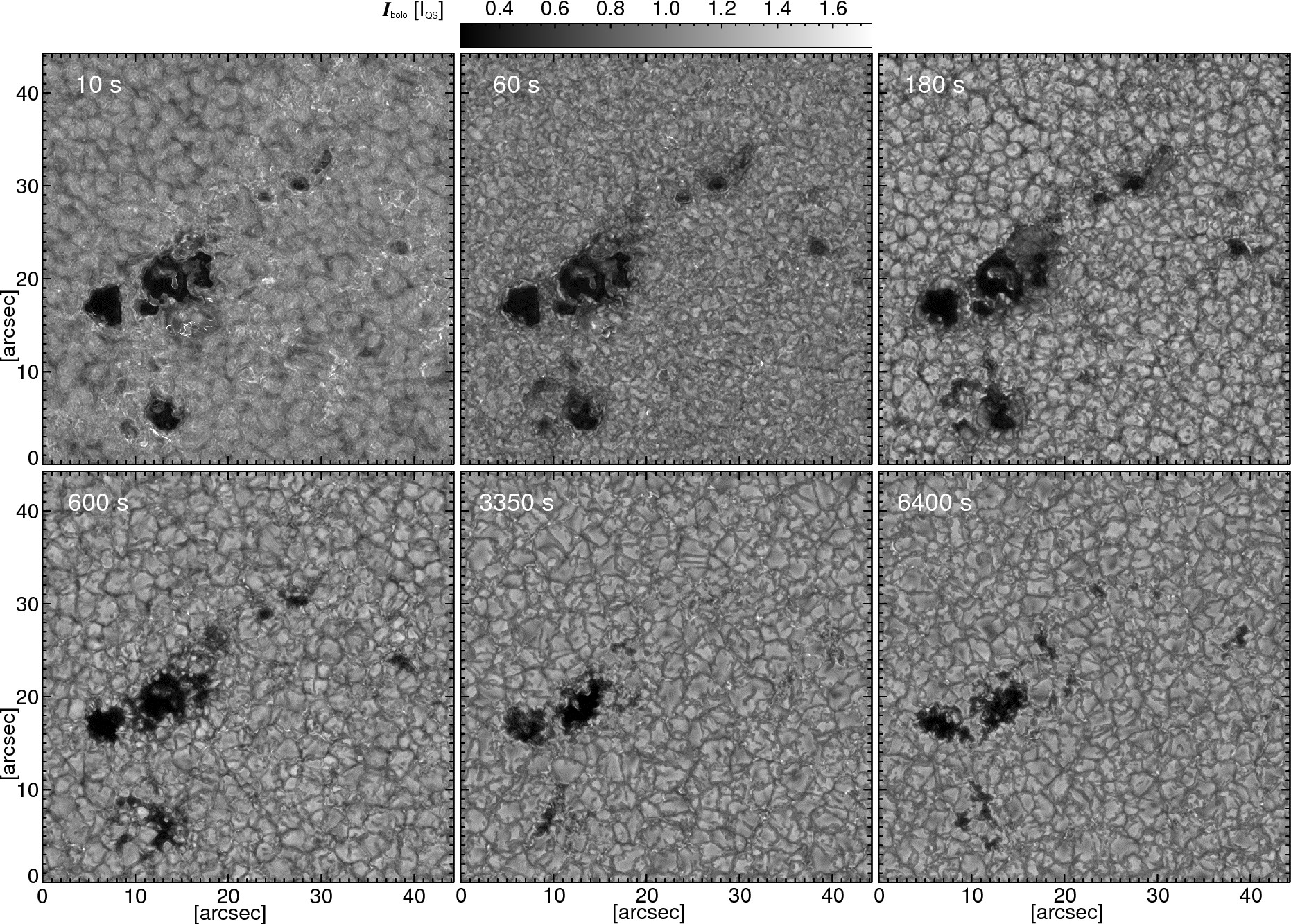}
\caption{Time evolution of the bolometric intensity of a MURaM simulation using the MASI first step results as inital condition. We present
snapshots at $t=10$\,s, 60\,s, 180\,s, 600\,s, 3350\,s, and 6400\,s (see text labels). An animation of this figure, together with the vertical
components of the velocity and magnetic field vector at optical depth unity, are provided with the online material.}
\label{Fig9}
\end{figure*}

Fig.~\ref{Fig9} displays maps of the bolometric intensity for snapshots taken at 10\,s, 1\,min, 3\,min, 10\,min, 56\,min, and 107\,min
of solar time after the start of the MURaM continuation (see text labels). The re-sorting of the MHD data by the MASI technique destroys
important physical properties of the system, e.g. the horizontal flow pattern is arbitrarily re-sorted, the horizontal
pressure balance is disturbed, and also the magnetic field lines are changed by the re-sorting, so that Maxwell's equation
$\nabla \cdot \vec{B} = 0$ is violated.

Even if the granulation pattern observed by IMaX is given as the initial condition, immediately after the start of the MURaM continuation
the IMaX granulation pattern gets destroyed, mainly caused by the unrealistic horizontal velocites. After about 3\,min of solar time a new
granulation pattern has evolved that is statistically similar, but looks different in detail from the IMaX granulation. The newly developed
granulation pattern indicates that after approximately 3\,min a new horizontal flow pattern has been established, at least in the upper layers
of the convection zone.

The MURaM code is also able to reestablish the physical consistency with respect to the magnetic field. Each MURaM calculation
can lead to tiny deviations from the $\nabla \cdot \vec{B} = 0$ constraint due to numerical rounding errors. Hence the code includes a
$\nabla \cdot \vec{B}$ cleaning algorithm which is of great benefit for our MURaM continuation of a re-sorted MHD data set. An analysis of the
temporal evolution of the $\nabla \cdot \vec{B}$~field (not shown) reveals that after 34\,min the value has reached the normal value seen before
the re-sorting.

Because on average pores live much longer than granules, their details also evolve within a few minutes but their rough shape remains unaffected
for longer time. Since for stray-light analyses it might be important to work with synthetic pores of the same size as in the observations,
it is interesting to mention that smaller pores (e.g. at the positions (24\arcsec,29\arcsec), (28\arcsec,30\arcsec), and (38\arcsec,23\arcsec)
in the right panel of Fig.~\ref{Fig2}) can disappear after a while (see snapshot $t=3350$\,s of Fig.~\ref{Fig9}), but because the magnetic flux
is conserved, pores of similar sizes can reappear at similar locations (see snapshot $t=6400$\,s).

An animation of the bolometric intensity as well as the vertical components of the velocity and magnetic field vector at optical depth unity
is available in the online edition.

\subsection{Iterative procedure}\label{Iterating}

Our considerations in the previous Sects.~\ref{Concept}-\ref{MuramContinuation} started with MHD data containing a solar scene that is similar to
the observed scene only in the sense that it contains similar types of features. After the MHD data were degraded, they could be used for a Stokes
inversion of the observation via the described method. The first iteration MASI result was a re-sorted MHD data set that was used as the initial condition of
a new MURaM simulation. After about half an hour the new simulation reached a statistically relaxed state in which the horizontal flow pattern and the
magnetic field lines were again in accordance with the MHD equations. That way we obtain MHD data containing a solar scene that is both, physically
consistent (at least in the visible layers of the Sun), and much more similar to our IMaX observations than the MHD data we started with. We now pursue
the question whether an iteration of the outlined method is able to improve the correspondance of the simulated and the observed data.

The reapplication of the method described in the following text is named the second MASI iteration. We use the snapshot at $t=600$\,s (bottom
left panel of Fig.~\ref{Fig9}) as the MHD data set for the second iteration because we aim for a close resemblance to the IMaX observation.
In particular we are interested in the smaller pores that are difficult to treat with the traditional Stokes inversion technique. Actually
one should only use snapshots taken after the relaxation process is entirely completed ($t \approx 34$\,min), but because $\nabla \cdot \vec{B}$
drops very rapidly during the first 10\,min after which it continues to slowly reach the value before the re-sorting after a further 24\,min,
the selected snapshot at $t=10$\,min strikes a reasonable balance. (We note that the smaller pores temporarily disappear for $t>10$\,min.)
The trick of doubling the MHD data set in order to make its magnetic field bipolar is no longer needed because the snapshot at $t=600$\,s
is already bipolar. 30 hours of execution time were needed to run the MURaM code for 600\,s of solar time on a cluster of 160 Intel(R)
Xeon(R) cores E5-2670.

\begin{figure}
\centering
\includegraphics[width=\linewidth]{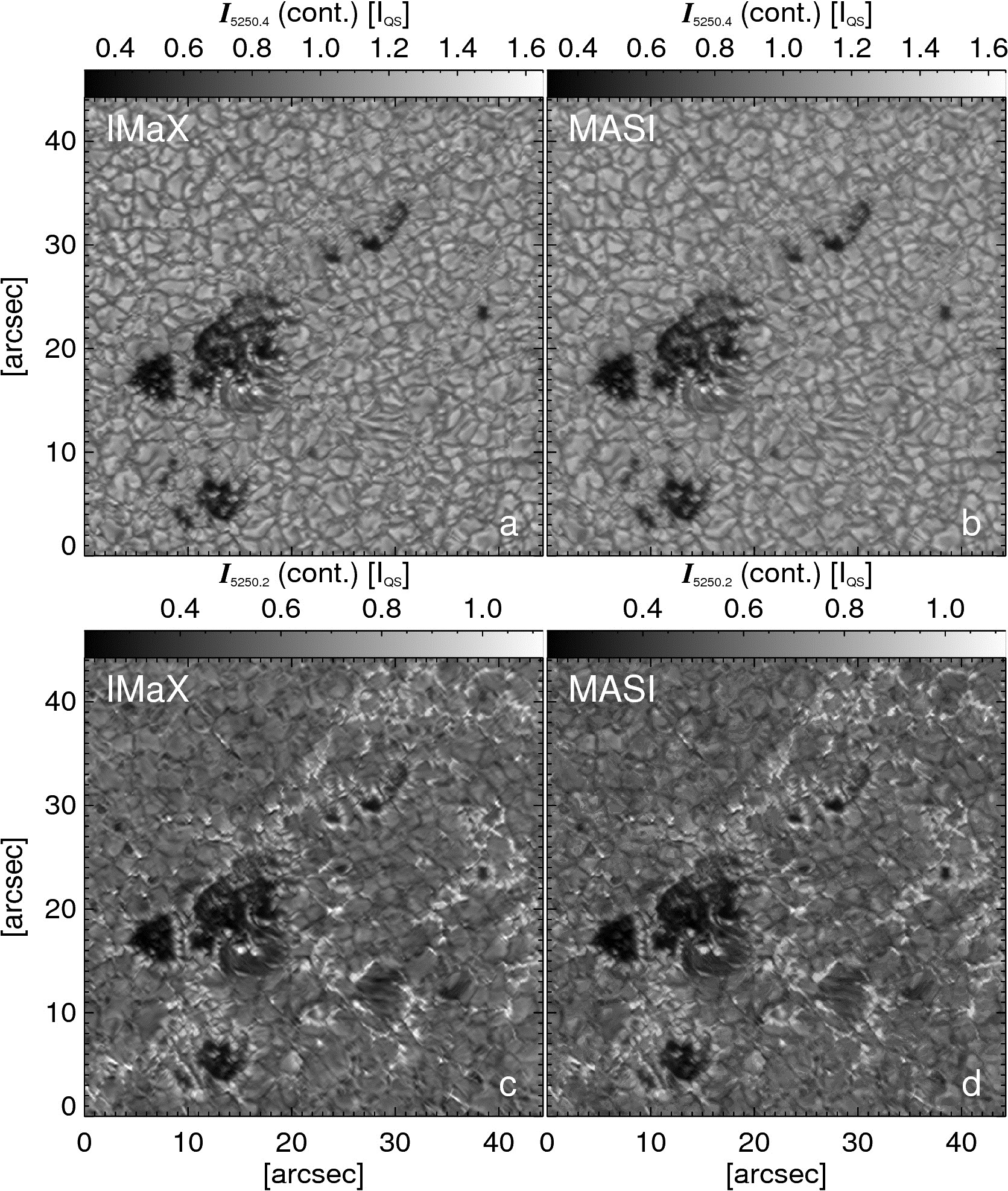}
\caption{Stokes~$I$ continuum (panel~a) and line-core (panel~c) image of the reconstructed IMaX observation and the degraded Stokes~$I$ continuum (panel~b)
and line-core (panel~d) image from the $2^{\mathrm{nd}}$ iteration MASI result.}
\label{Fig10}
\end{figure}

A further change in the second iteration is that we this time invert reconstructed IMaX data (level-2) via MASI in order to demonstrate that
the MASI method leaves the user the choice, whether the observations are deconvolved with a PSF or the MHD data are convolved. This can be
an advantage, because a deconvolution of observational data with a PSF leads regularly to a decrease in the signal-to-noise ratio, while it
provides a higher spatial resolution (compare panel~a of Fig.~\ref{Fig10} with the right panel of Fig.~\ref{Fig2}). As a
consequence, the convolution of the MHD data with the PD PSF is skipped this time because the PD PSF is already considered during the reconstruction
of the IMaX data. All other degradation steps (convolution with the spectral PSF, Butterworth lowpass filtering, and 25\,\% global stray-light contamination)
have to be applied in the same way as for the first iteration. The RMS quiet-Sun intensity contrast of the degraded synthetic Stokes~$I$ continuum image, 13.67\,\%,
agrees very well with the contrast of the IMaX level-2 data, 13.25\,\%, likewise the intensity of the darkest pixel ($0.308~I_{\rm{QS}}$ for the synthetic data
and $0.285~I_{\rm{QS}}$ for the IMaX data).

The similarity between simulation and observation makes it possible to extend the merit
function by a term that slightly reduces the demolition of the horizontal flow pattern and magnetic field lines by the re-sorting:
   \begin{equation}\label{Eq_Merit2}
   \chi^2=C D + \frac{1}{32} \sum_{w=1}^{8}{ \sum_{s=1}^{4}{\frac{(I_{w,s}^{\rm{obs}}(x_o,y_o)-I_{w,s}^{\rm{syn}}(x_m,y_m))^2}{\sigma_s^2}}}~,
   \end{equation}
where $C$ is a constant that determines the strength of the correction term and
   \begin{equation}\label{Eq_Dist}
   D=\sqrt{(x_o-x_m)^2+(y_o-y_m)^2}
   \end{equation}
is the distance between the considered observed pixel and the synthetic one. The correction term,
$C D$, gives a preference to synthetic pixels that are located close to the observed pixel. In our example we set $C=0.004~\mathrm{km}^{-1}$, which is
a relatively small value, in order to prefer neighbouring pixels only in the case that multiple synthetic pixels possess almost identical Stokes profiles.
With this selection, 19\,\% of the $812 \times 812$ entries in the archive of the second iteration provide the best-fit results used for
the re-sorted data set.

\begin{figure}
\centering
\includegraphics[width=\linewidth]{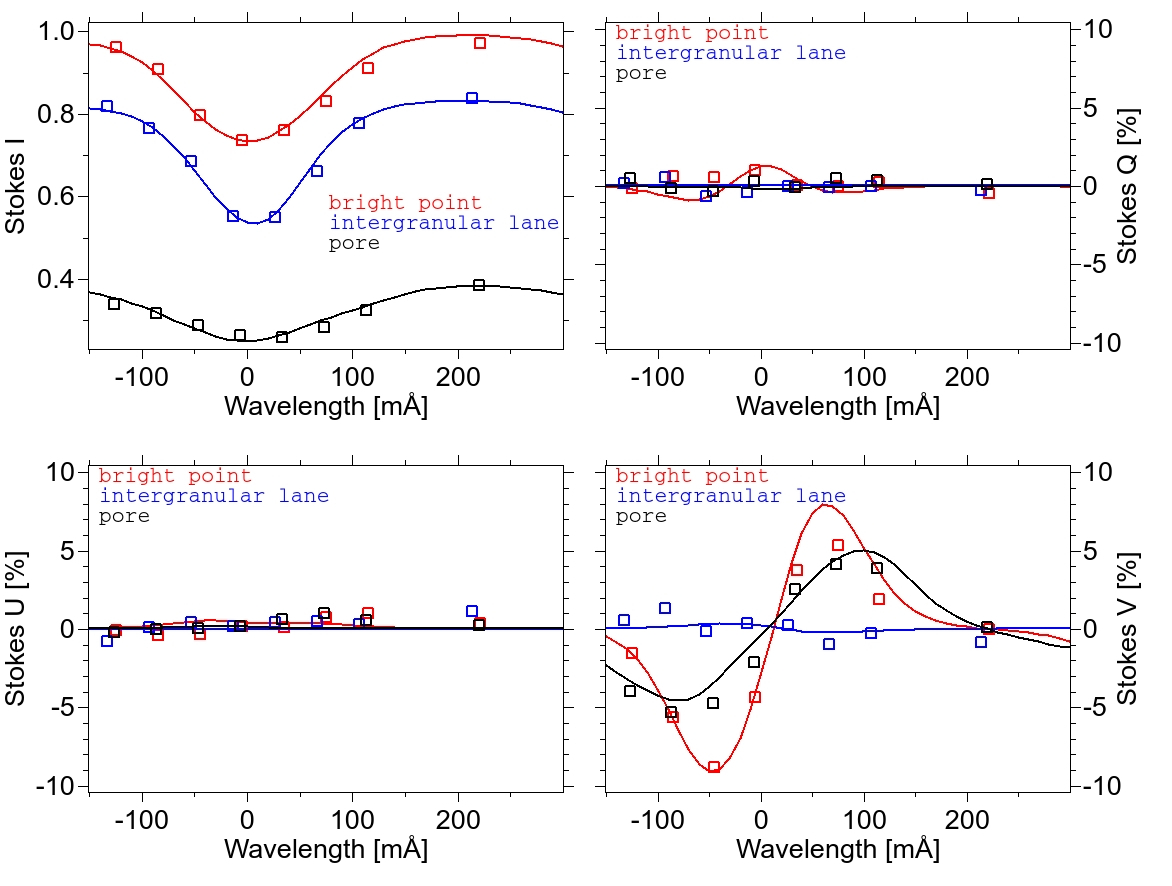}
\caption{Stokes $I$, $Q$, $U$, and $V$ profiles from a position inside a bright point (red), an intergranular lane (blue), and a pore (black).
The squares display the reconstructed and blueshift-corrected IMaX observation, while the solid lines correspond to the $2^{\mathrm{nd}}$ MASI iteration,
i.e. the best-fit inversion results. All profiles are normalized to the spatially averaged quiet-Sun intensity.}
\label{Fig11}
\end{figure}

Fig.~\ref{Fig10} contrasts the Stokes~$I$ continuum and line-core image of the reconstructed IMaX observation with the degraded MASI result of the second iteration.
Compared to the first iteration the agreement has slightly improved. Even the penumbra-like region does not show any obvious artifacts. A comparison
of the observed Stokes profiles (reconstructed and corrected for the etalon blueshift) with the best-fit profiles resulted from the second MASI iteration can
be seen in Fig.~\ref{Fig11}. We selected three pixels located at the positions (13\carcsec{}89,26\carcsec{}79), (17\carcsec{}48,39\carcsec{}37),
and (6\carcsec{}43,18\carcsec{}35) in Fig.~\ref{Fig10} and representing a bright point, an intergranular lane, and a pore. MASI provides for the three pixels
a magnetic field strength at $\log(\tau)=-1$ of 1520\,G, 205\,G, and 2490\,G, respectively. The kilogauss field in the pore and in the bright point broadens the
spectral line. The low temperature gradient in the pore together with the temperature sensitivity of the Fe\,{\sc i} 5250.2\,\AA{} line leads to a shallow
Stokes~$I$ profile. Since the magnetic field is weak for the intergranular-lane pixel, the corresponding polarization signals are also weak.
The magnetic field of the pixel in the pore and the pixel in the bright point is strong and nearly vertical so that the Stokes~$Q$ and $U$ signals are weak,
while the Stokes~$V$ profiles reach values of a few percent.

\begin{figure}
\centering
\includegraphics[width=\linewidth]{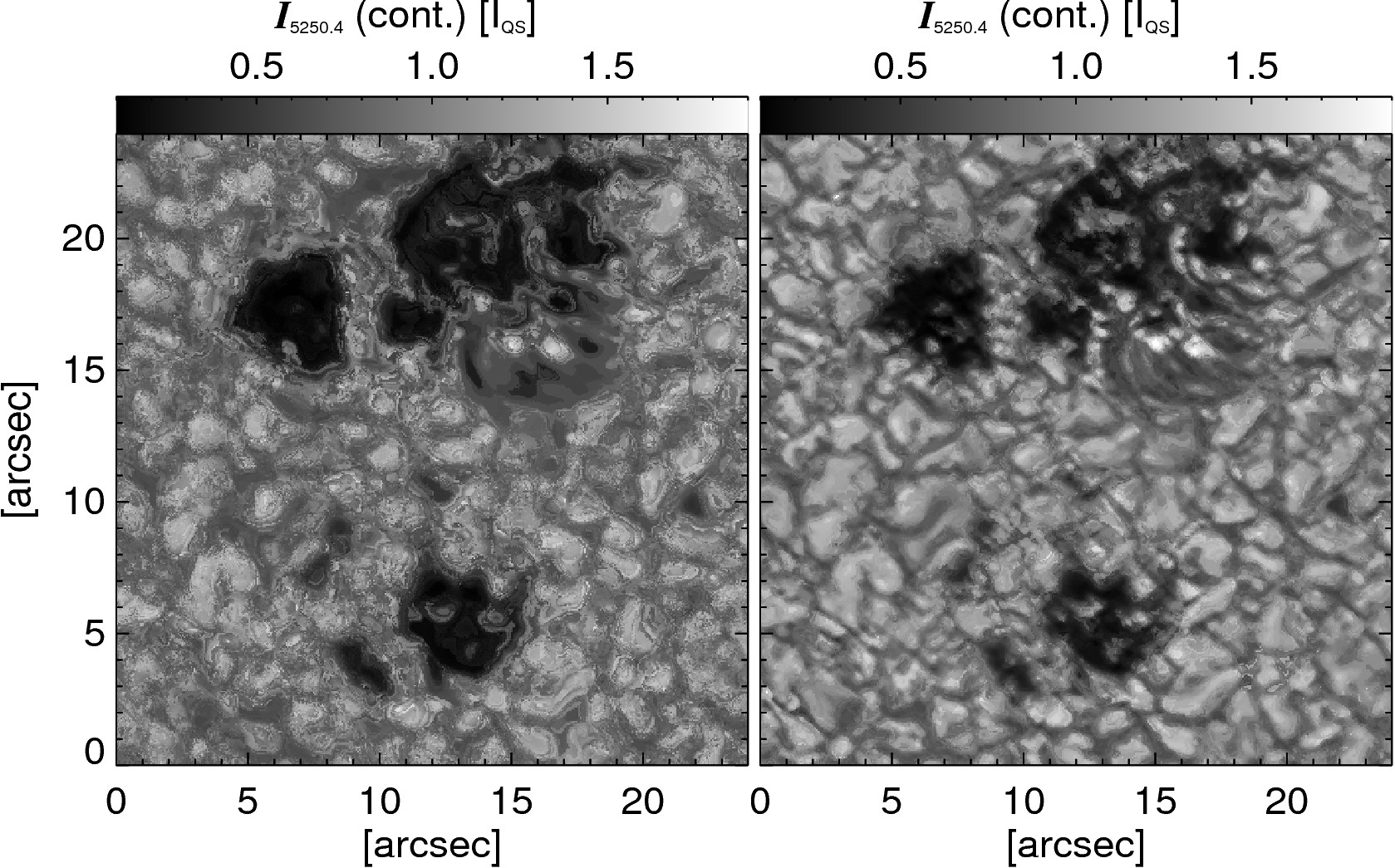}
\caption{Undegraded Stokes~$I$ continuum image after the first (left panel) and second (right panel) MASI iteration.
Only the bottom left $24\arcsec{} \times 24\arcsec{}$ part of the considered field of view is shown for a better visibility of details.}
\label{Fig12}
\end{figure}

A look at the undegraded Stokes~$I$ continuum image of the re-sorted MHD data set demonstrates the improvements of the second iteration even more distinctly.
While the result of the first iteration showed a wealth of small-scale discontinuities (see the left panel of Fig.~\ref{Fig12}), this is rarely the case for the ouput of the second
iteration (right panel of Fig.~\ref{Fig12}).

\begin{figure}
\centering
\includegraphics[width=\linewidth]{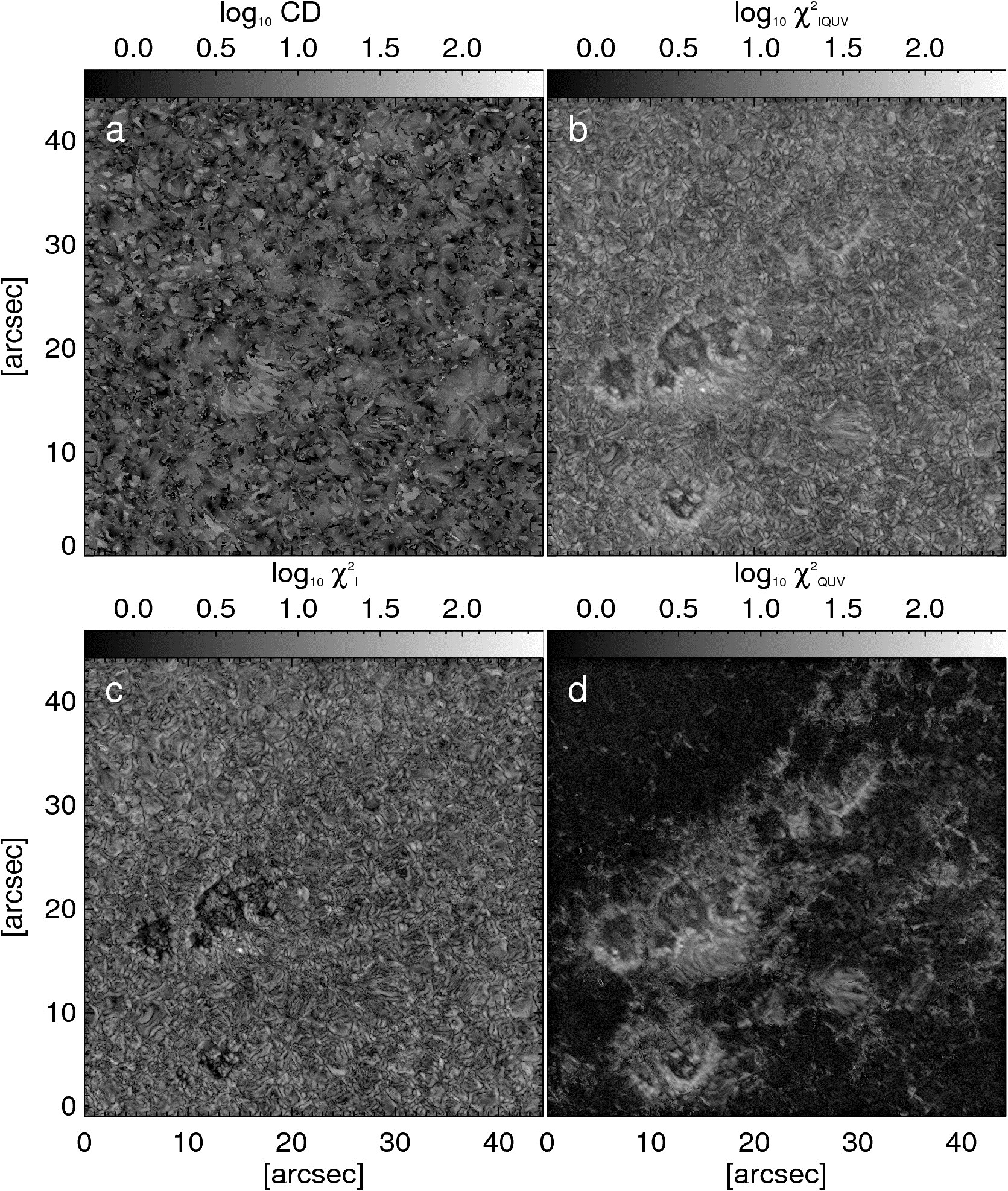}
\caption{Contributions to the total $\chi^2$ values of the second iteration provided by the distance term
in Eq.~\ref{Eq_Merit2} (panel~a), by the sum of all four Stokes parameters (panel~b), by only Stokes~$I$ (panel~c), and by the sum of Stokes~$Q$,
$U$, $V$ (panel~d). All panels are plotted on a logarithmic scale.}
\label{Fig13}
\end{figure}

The improvement of the inversion's quality can also be seen by a comparison of the $\chi^2$ maps. Panels~a and b of Fig.~\ref{Fig13} show
the total $\chi^2$ map of the second iteration decomposed into the distance term, $C D$, and the remaining Stokes part, $\chi^2_{IQUV}$.
The Stokes part is further decomposed into the contribution of purely Stokes~$I$ ($\chi^2_{I}$, panel c of Fig.~\ref{Fig13}) and the
remaining Stokes parameters ($\chi^2_{QUV}$, panel d), which allows a direct comparison of the $\chi^2$ components of the first iteration
(panels b-d of Fig.~\ref{Fig8}) with the ones of the second iteration (panels b-d of Fig.~\ref{Fig13}). The maximum $\chi^2_{IQUV}$~value
of the second iteration is $1.9$~times lower than for the first iteration. The mean $\chi^2_{IQUV}$~value decreased from $17.1$ to $7.9$,
the mean $\chi^2_{I}$~value from $10.9$ to $5.8$, and the mean $\chi^2_{QUV}$~value from $6.3$ to $2.1$. On average, the additional
correction term, $C D$, contributes 34\,\% to the total $\chi^2$~value, while 48\,\% originate from Stokes~$I$, and 18\,\% from Stokes~$Q$, $U$,
and $V$ together.

\begin{figure}
\centering
\includegraphics[width=\linewidth]{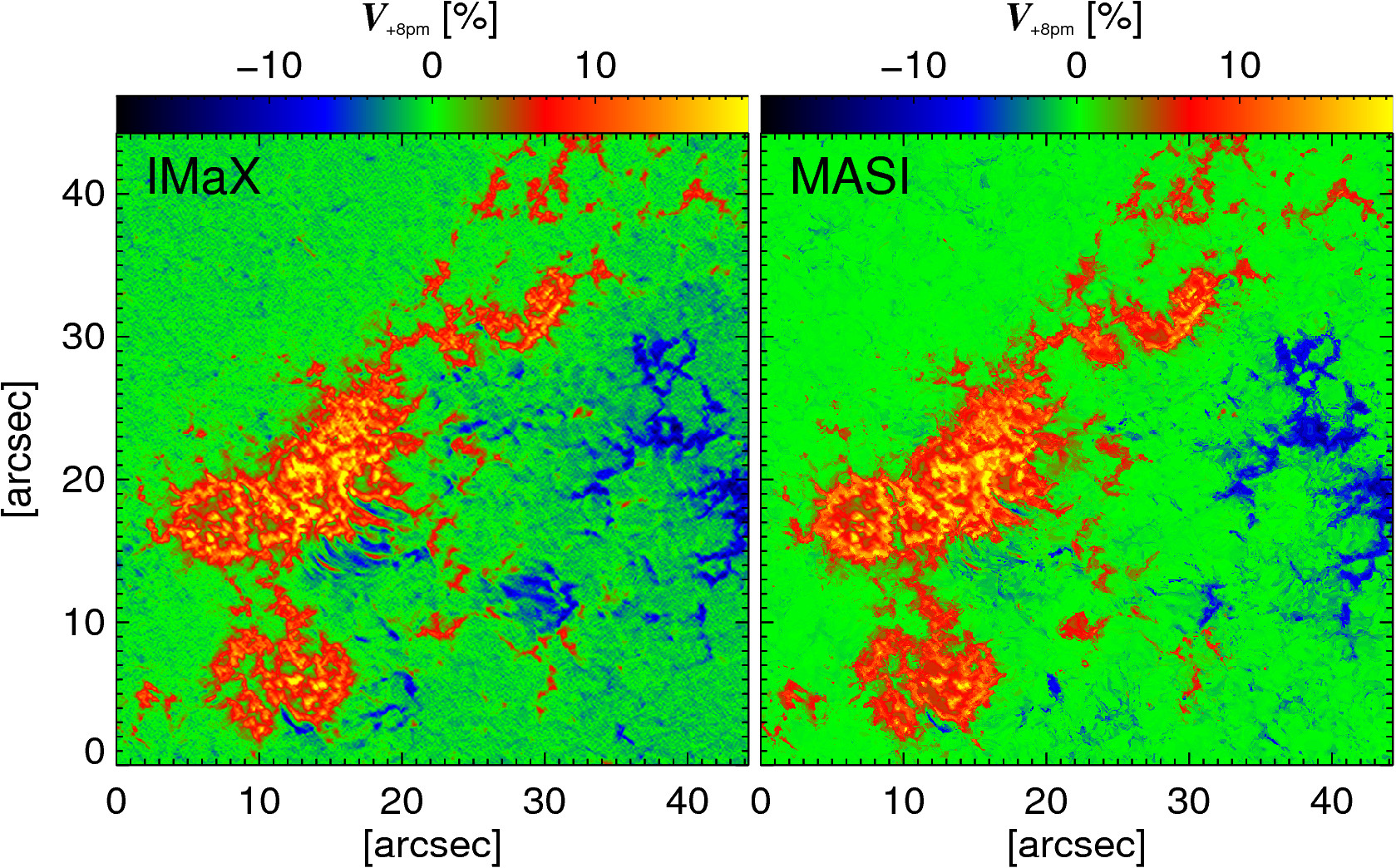}
\caption{Normalized Stokes~$V$ image at $+8$~pm offset from the line core. Left panel: Reconstructed IMaX data. Right panel: Degraded
MHD data re-sorted by the second MASI iteration.}
\label{Fig14}
\end{figure}

Stokes~$V$ maps in the red wing of the 525.02\,nm line are displayed for the reconstructed IMaX observation and the degraded MASI
result in Fig.~\ref{Fig14}. The reconstructed IMaX Stokes~$V$ image shows much more fine structure than the level-1 data (see left
panel in Fig.~\ref{Fig7}). Most of the fine structure is well reproduced by the second MASI iteration, although some
features are not well fitted, largely the ones with a nearly horizontal magnetic field. Nonetheless, the discrepancy between observation
and MASI at the position of the penumbra-like feature and also of the flux-emergence region is lower than for the first
MASI iteration.

\begin{figure}
\centering
\includegraphics[width=\linewidth]{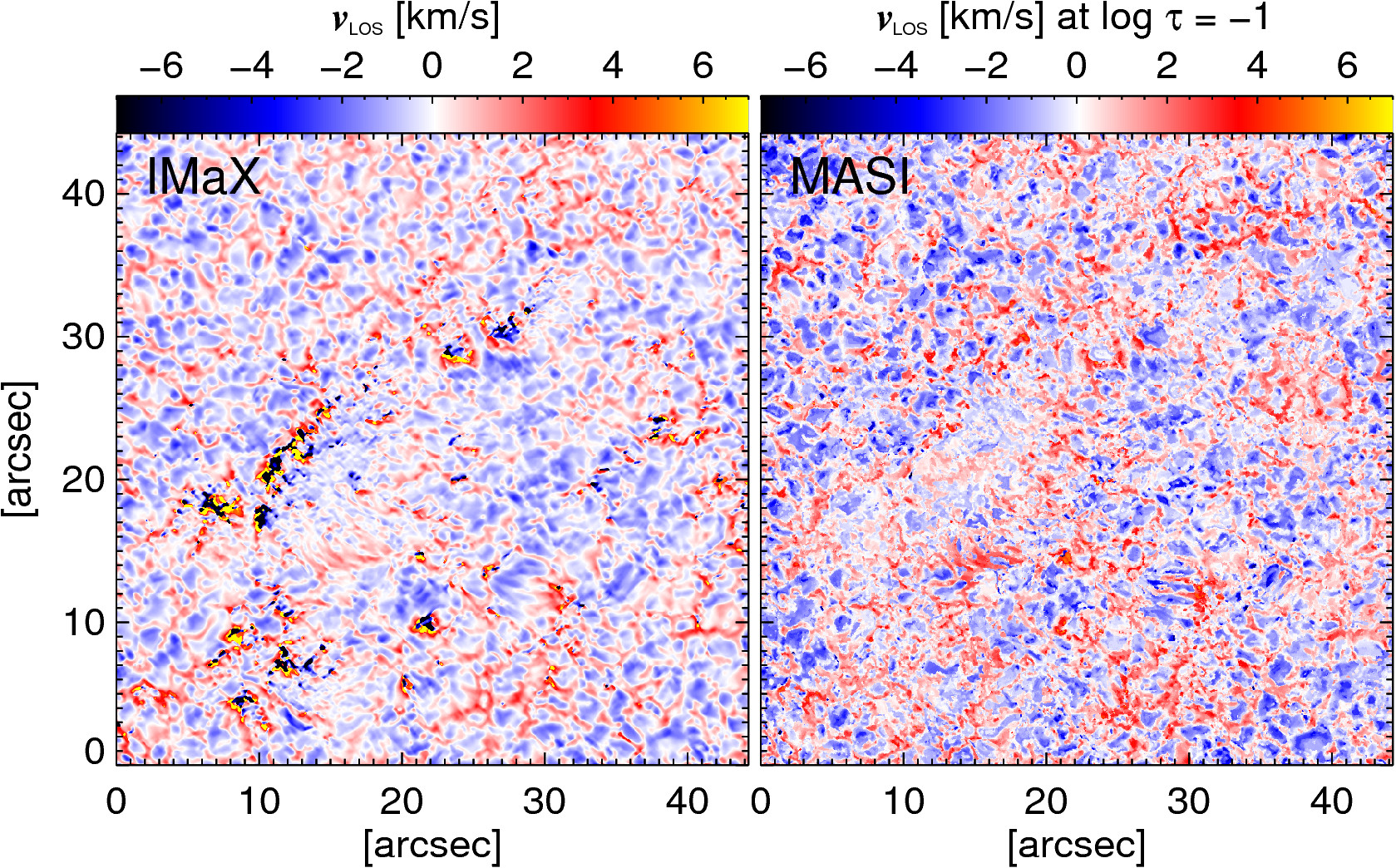}
\caption{Map of the line-of-sight velocity. Left panel: Classical estimate from the reconstructed data observed by IMaX (Stokes~$I$ Gaussian fit).
Right panel: Undegraded MHD data at $\log(\tau)=-1$ re-sorted by the second MASI iteration. Negative velocities correspond to upflows.}
\label{Fig15}
\end{figure}

\begin{figure}
\centering
\includegraphics[width=\linewidth]{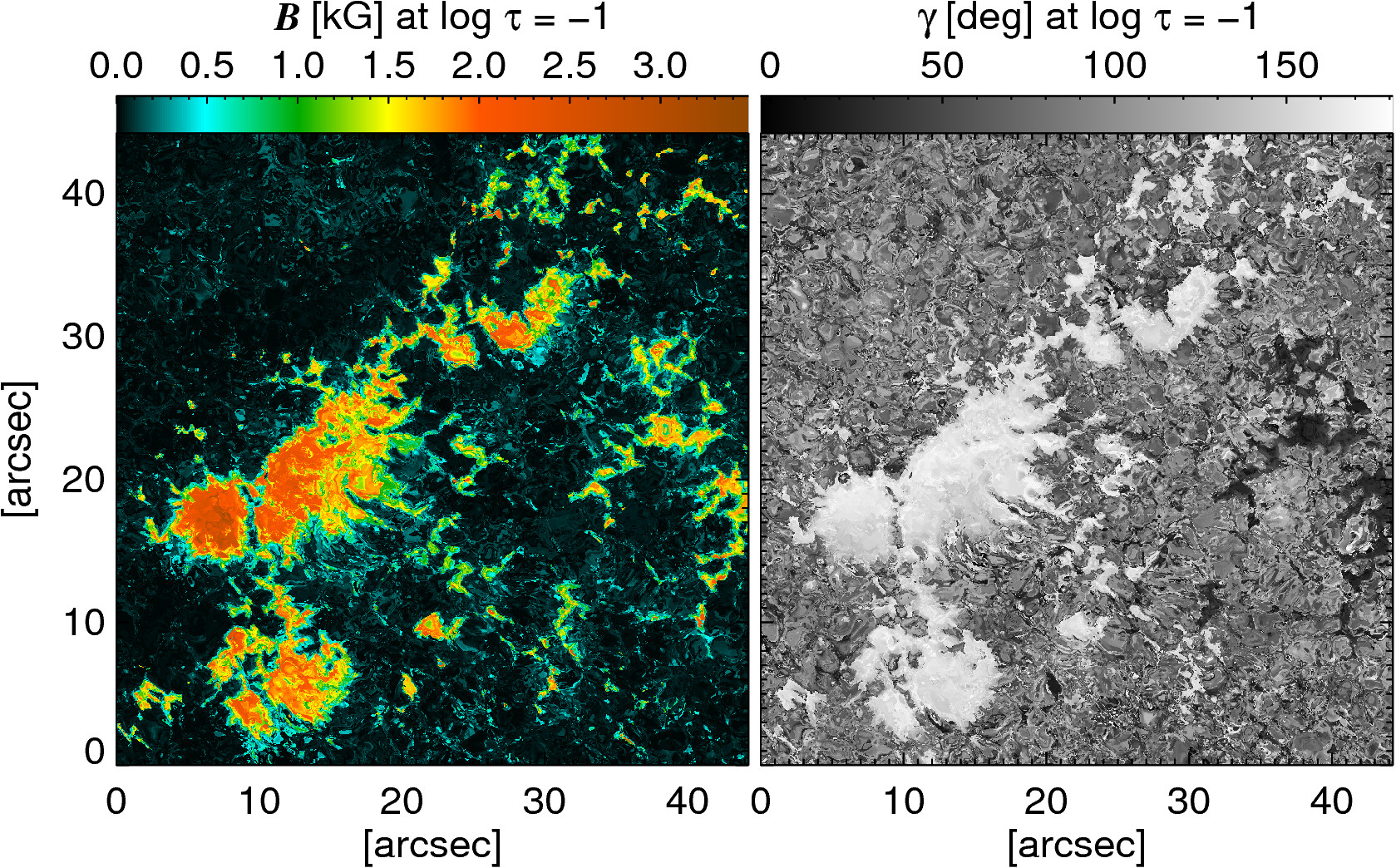}
\caption{Magnetic field strength (left panel) and inclination (right panel) at $\log(\tau)=-1$, both taken from the undegraded second iteration MASI result.}
\label{Fig16}
\end{figure}

We consider some more undegraded quantities of the second MASI iteration in Figs.~\ref{Fig15}~to~\ref{Fig16}.

Fig.~\ref{Fig15} compares maps of the LOS velocity between IMaX and MASI. Since a 25\,\% global stray-light contamination is part of the MASI degradation, we
applied a corresponding stray-light correction to the reconstructed IMaX data and fitted the Stokes~$I$ profiles with a Gaussian. The central position
of the Gaussian (corrected for the blueshift) provides the LOS velocities in the left panel of Fig.~\ref{Fig15}. The right panel displays the vertical
component of the re-sorted MHD velocity vector at the optical depth $\log(\tau)=-1$, because this depth is roughly the formation height of the 525.02\,nm line.
The velocity scale was shifted to reach a zero mean value over the entire FOV.

For both panels of Fig.~\ref{Fig15} the granulation outside the pores exhibits the typical upflows within granules and downflows for the intergranular lanes.
The contrast of the MASI result is larger than the one of the observation, possibly because the formation height of the 525.02\,nm is somewhat higher
than $\log(\tau)=-1$. For display reasons we limited the velocity range to $\pm 7$\,km/s because the Gaussian fits of the observed profiles have significant
difficulties at the edges and inside the pores (see the black and yellow regions in the left panel of Fig.~\ref{Fig15}). In the pores the temperatures are low
and the temperature gradiants are small, both leading to shallow spectral lines. In combination with the photon noise the Gaussian fits can lead to unreasonable
results.

Finally, we present the magnetic field of the second MASI iteration. The left panel of Fig.~\ref{Fig16} displays the field strength, while the right panel
shows the field inclination, both at the optical depth of $\log(\tau)=-1$. In the pores we find field strengths up to 3400\,G. We note that the MASI
results hardly show any field weakening in the pores, that we mentioned in Sect.~\ref{StrayLight}, which further indicates that our simplistic stray-light model
is not too far off from reality.

\section{Summary and discussion}\label{Summary}

We obtained atmospheric parameters such as temperature, magnetic field vector, and LOS velocity from a spectropolarimetric observation by finding
the best matched Stokes profiles from a snapshot of an MHD simulation, degraded as realistically as possible to the level of the observed data.

Because the MHD data set contains the full set of atmospheric parameters, we were able to re-sort the vertical columns of the original MHD data
corresponding to the best-fits of observational Stokes profiles and used the re-sorted data as the initial condition of a new MHD simulation.
Since the spatial relations of the columns are destroyed (in the horizontal directions) by the re-sorting, the initial condition is
physically inconsistent. The inconsistencies include the re-arranged horizontal flow pattern both, above the solar surface
and below it, and the loss of local horizontal pressure balance, which are
closely related to the convection in the granulation cells, so that the observed granulation pattern gets destroyed immediately after the start of
the MURaM continuation. The MURaM code is able to remove the physical inconsistencies, so that after about 3\,min a new granulation pattern has
evolved and after about half an hour of solar time the divergence of the magnetic field vector is brought down to zero. Further studies are needed
to find ways to reduce the relaxation time, e.g. by carrying out more iterations or by advancing the applied merit function to preserve the horizontal
flows of a granule in a statistical way (as we tried by adding the very simplistic distance term in Eq.~\ref{Eq_Merit2}).

After the relaxation we obtained physically consistent MHD simulations that show a solar scene quite similar to the observed one.
We used a snapshot of the new simulations to apply a second iteration of the method. The higher similarity to the observation improved the match
between observed and synthetic Stokes profiles. In particular, the improvement can be seen if undegraded Stokes images of the two iterations are compared
and also by a significant decrease in the mean $\chi^2$ value and its contributions from Stokes~$I$ and from Stokes~$Q$, $U$, $V$, respectively.

The high computational effort needed for the MURaM simulations hindered the realization of further
iterations within this study, which are expected to lead to a stepwise convergence between observation and simulation, although it is clear that
effects of an imperfect modelling of instrumental effects and imperfections of the MHD code (e.g., for penumbrae) cannot be lowered by more iterations.

The MASI technique can be understood as a first step towards an integration of the MHD equations into the Stokes inversion of a time series.
Although we are still a long way away from this long-term goal, a few applications of the technique can already be considered:

\begin{enumerate}
   \item By means of the MASI technique we are able to create MHD simulations that are very similar to an observation. This can be quite helpful
   in analyzing physical phenomena. For example, the MHD data allow for insights into all physical quantaties, even the ones that are not covered by
   the spectral lines (horizontal velocities, densities, pressures, horizontal force balance). It can also be advantageous that the MHD data are free of noise and
   can have higher spectral, spatial, and temporal resolution. Moreover, the availability of an MHD simulation similar to the observation can be
   also quite useful in analyzing instrumental effects (e.g., the stray-light behavior).
   \item The MASI method allows the creation of new MHD simulations with interesting solar targets, e.g., light bridges, complex sunspot or pore topologies,
   which was difficult in the past.
   \item The MASI results after the first step can be used as a first-guess atmosphere for a traditional Stokes inversion technique so that the traditional
   inversion can converge faster.
\end{enumerate}

Although the technique presented here is promising, this paper is limited to first steps and much work remains to be done to test and improve it further.
A first important test could be a comparison between the MASI results and the results of traditional inversion techniques. Ideally, this should be done with realistically
degraded MHD data, because then the errors caused by the MASI technique can be disentangled from the errors made by traditional inversions, which would not be easily possible
if observational data are inverted. Furthermore, it should be investigated if more iterations lead to further improvements and if so how many iterations are needed until
the $\chi^2$ values converge to its final value. An application of the MASI method to other observational data sets is also desirable, in particular to data with
full line profiles (i.e. better spectral sampling) and a better knowledge of instrumental effects. Additionally, the MHD archive should be expanded by features harboring more
horizontal fields (flux emergence, penumbrae) and local-dynamo simulations. Possibly, the performance of MASI improves if the re-sorted MHD data are spatially smoothed
\citep[similar to the smoothing applied as part of the SPINOR inversion of the IMaX data, see][]{Solanki2016} before being used as initial condition of an MHD simulation. It also appears advisable
to introduce Stokes-specific weighting factors in the merit function in order to reach a better balance between the contributions of the four Stokes parameters. Finally,
an error estimate of the method should be derived, e.g., by considering all vertical columns of the original MHD data set that lead to very similar $\chi^2$ values.

\begin{acknowledgements}
The German contribution to \sunrise{} and its reflight was funded by the
Max Planck Foundation, the Strategic Innovations Fund of the President of the
Max Planck Society (MPG), DLR, and private donations by supporting members of
the Max Planck Society, which is gratefully acknowledged. The Spanish
contribution was funded by the Ministerio de Econom\'ia y Competitividad under
Projects ESP2013-47349-C6 and ESP2014-56169-C6, partially using European FEDER
funds. The National Center for Atmospheric Research is sponsored by the National
Science Foundation. The HAO contribution was partly funded through NASA grant number
NNX13AE95G. The National Solar Observatory (NSO) is operated by the Association of
Universities for Research in Astronomy (AURA) Inc. under a cooperative agreement
with the National Science Foundation. This work was partly supported by the BK21
plus program through the National Research Foundation (NRF) funded by the Ministry
of Education of Korea.
\end{acknowledgements}

\end{document}